%% file: arXiv_2024_log_parsing.tex
\PassOptionsToPackage{dvipsnames,table}{xcolor}
\documentclass[sigconf]{acmart}

\usepackage{multirow}
\usepackage{subfig}
\usepackage{xcolor,colortbl}
\definecolor{lightgray}{gray}{0.9}

\usepackage{listings}
\newcolumntype{a}{>{\columncolor{lightgray}}r}
\newcolumntype{b}{>{\columncolor{white}}r}

\usepackage{fancybox}

\usepackage{framed}   %

\newenvironment{myframed}{%
  \MakeFramed {\advance\hsize-\width \FrameRestore}}%
 {\endMakeFramed}

\newcommand{\head}[1]{\par\noindent\textbf{#1:}\space}

\copyrightyear{2024}
\acmYear{2024}
\setcopyright{rightsretained}
\acmConference[ESEM '24]{Proceedings of the 18th ACM / IEEE International Symposium on Empirical Software Engineering and Measurement}{October 24--25, 2024}{Barcelona, Spain}
\acmBooktitle{Proceedings of the 18th ACM / IEEE International Symposium on Empirical Software Engineering and Measurement (ESEM '24), October 24--25, 2024, Barcelona, Spain}
\acmDOI{10.1145/3674805.3686684}
\acmISBN{979-8-4007-1047-6/24/10}

\makeatletter
\def\@copyrightpermission{
  \hspace*{0mm}\includegraphics[width=2cm]{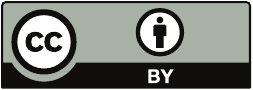}%
  \hspace*{2mm}\raisebox{2.5mm}[25pt][5pt]{%
          \parbox{\columnwidth}{\footnotesize This work is licensed under a Creative Commons \\ Attribution 4.0 International (CC BY 4.0) license.}%
  }%
}%
\makeatother

\begin{document}
\title{A Comparative Study on Large Language Models for Log Parsing}

\author{Merve Astekin}
\email{merve@simula.no}
\affiliation{
  \institution{Simula Research Laboratory}
  \city{Oslo}
  \country{Norway}
}

\author{Max Hort}
\email{maxh@simula.no}
\affiliation{
  \institution{Simula Research Laboratory}
  \city{Oslo}
  \country{Norway}
}

\author{Leon Moonen}
\email{leon.moonen@computer.org} %
\affiliation{
  \institution{Simula Research Laboratory \& \newline 
               BI Norwegian Business School}
  \city{Oslo}
  \country{Norway\\[1.5ex]}
}

\begin{abstract}

\textit{Background}: Log messages provide valuable information about the status of software systems.
This information is provided in an unstructured fashion and automated approaches are applied to extract relevant parameters. 
To ease this process, log parsing can be applied, which transforms log messages into structured log templates.
Recent advances in language models have led to several studies that apply ChatGPT to the task of log parsing with promising results. 
However, the performance of other state-of-the-art large language models (LLMs) on the log parsing task remains unclear. 

\textit{Aims}: In this study, we investigate the current capability of state-of-the-art LLMs to perform log parsing.

\textit{Method}: We select six recent LLMs, including both paid proprietary (GPT-3.5, Claude 2.1) and four free-to-use open models, and compare their performance on system logs obtained from a selection of mature open-source projects. 
We design two different prompting approaches and apply the LLMs on $1,354$ log templates across 16 different projects.
We evaluate their effectiveness, in the number of correctly identified templates, and the syntactic similarity between the generated templates and the ground truth. %

\textit{Results}: We found that free-to-use models are able to compete with paid models, with CodeLlama extracting 10\% more log templates correctly than GPT-3.5.
Moreover, we provide qualitative insights into the usability of language models (e.g., how easy it is to use their responses).

\textit{Conclusions}: Our results
reveal that some of the smaller, free-to-use LLMs can considerably
assist log parsing compared to their paid proprietary competitors,
especially code-specialized models. 

\end{abstract}



\maketitle

\section{Introduction}

Software systems produce system logs that offer valuable information about the system's status and execution. 
Developers and operators commonly leverage these logs to ensure the reliability of software systems, as they represent the only source of data that records software runtime information.
This information serves as a foundation for fault detection, prediction, diagnosis, and recovery activities.

\begin{figure*}
\includegraphics[width=.85\linewidth]{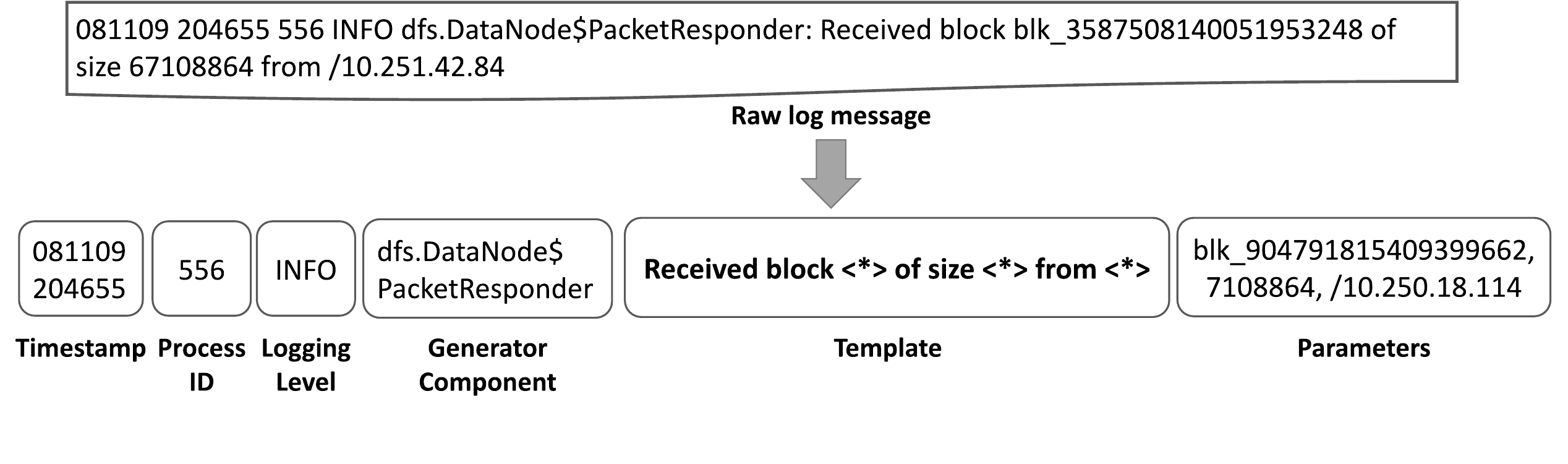}
\vspace*{-2ex}\caption{An example workflow of log parsing.}
\label{fig:log-parsing}
\end{figure*}

Log messages consist of natural text as well as dynamic variables that are determined at the time of writing the message. 
An example log message is shown in Figure~\ref{fig:log-parsing}.
To make use of such a message, one first needs to transform the data into a structured format via the application of log parsers.

Two main approaches have been followed for parsing log data~\cite{jiang2024:lilac}:
\textit{(i) Syntax-based approaches} that utilize pre-designed features and rules requiring domain knowledge; 
\textit{(ii) Semantic-based approaches} that usually employ deep neural networks to learn semantics within log messages.
Most of the existing research is directed towards syntax-based log parsing~\cite{he2017:drain, du2016:spell, astekin2019:dilaf, dai2022:logram}. 
However, it is known that syntax-based log parsers struggle when faced with log data in different formats that do not comply with predefined rules~\cite{zhu2019:tools}. 
As an alternative to syntax-based log parsers, semantic-based log parsing techniques have been proposed by several studies~\cite{huo2023:semparser, le2022:logbased, li2023:did, liu2022:uniparser}.
By extracting semantic information from log messages, semantic-based log parsers can increase parsing accuracy. 
However, recent benchmark studies~\cite{khan2022:guidelines, jiang2024:largescale} revealed insufficient performance on larger and more diverse log datasets.

The similarities between software system logs and natural language have directed researchers to explore language models for log analysis. 
As with other software engineering tasks~\cite{mudgal2023:assessment}, Large Language Models (LLMs) gained popularity for performing log-oriented tasks, such as log parsing~\cite{le2023:loga, xu2024:divlog, jiang2024:lilac}, and log anomaly detection~\cite{liu2024:interpretable, qi2023:loggpt}. 
Here, the majority of studies based on LLMs employ ChatGPT~\cite{jiang2024:lilac,le2023:log,mudgal2023:assessment}.
Although ChatGPT achieves promising results for log parsing, there is a lack of benchmarking studies that compare different LLMs to determine how and where they differ and find which ones are best suited for log parsing.
Moreover, from a privacy and cost perspective, it is unclear whether there exist free-to-use models that perform competitively with ChatGPT for the log parsing task.

To address this gap, we analyze the performance of six recent LLMs—including both proprietary and free-to-use models—on system logs collected from a range of widely used open-source projects. 
As part of our analysis, we conclude several observations that help interpret the capability of cutting-edge LLMs to perform log parsing across a wide range of system logs.
Moreover, these observations reveal that some of the smaller, free-to-use LLMs can considerably assist log parsing compared to their paid proprietary competitors. This is especially true for code-specialized models, such as CodeLlama~\cite{roziere2023:code}, which is the best performing LLM on two out of three investigated metrics.

\section{Related Work}
Automated log analysis techniques mostly require structured data. 
Log parsing, which aims to convert unstructured log messages into structured log templates and extract the parameters, is typically the initial step toward automated log analysis.
Although there have been several studies that aim at parsing system logs automatically, we focus on LLM-based approaches and evaluation metrics in this section.

\subsection{Log Parsing with Large Language Models}
\label{related:llms}
Pre-trained language models, or PTMs, have been applied in several studies for different phases of log analysis, including anomaly detection~\cite{guo2021:LogBERT, chen2022:bertlog, lee2023:lanobert} and log parsing~\cite{nedelkoski2021:selfsuperviseda, le2023:log}. 
Le and Zhang~\cite{le2023:log} proposed LogPPT, for which they trained a RoBERTa model with few-shot learning. 
The model receives a log message as input and for each token is asked to predict whether it is a dynamic parameter or not.

Several recent and concurrent studies adopt LLMs, such as ChatGPT, for various software engineering tasks~\cite{mudgal2023:assessment}, including log parsing~\cite{le2023:loga, xu2024:divlog, jiang2024:lilac}, and log anomaly detection~\cite{liu2024:interpretable, qi2023:loggpt}. 
Jiang et al.~\cite{jiang2024:lilac} presented the LILAC framework, which queries ChatGPT (gpt-3.5-turbo-0613), Davinci, and Curie for log parsing. 
LILAC applies a selection procedure to find relevant template extraction examples to add to the prompts.
Moreover, it stores the generated templates in a cache. 
For every new log message, the cache is queried to determine whether a corresponding log template has already been generated. 
This reduces the number of times LLMs have to be queried.
Xu et al.~\cite{xu2024:divlog} proposed the DivLog framework, which applies few-shot prompting with an optimized procedure for selecting examples for the prompt.

Ma et al.~\cite{ma2024:llmparser} proposed an LLM-based log parser, LLMParser, which employs four different LLMs: Flan-T5-small, Flan-T5-base, LLaMA-7B, and ChatGLM-6B. 
They investigated the effect of training size, model size, and pre-training on log parsing accuracy.
They experimented only with models pre-trained on natural language tasks. 

Le and Zhang~\cite{le2023:loga} studied the ability of ChatGPT for log parsing under consideration of different prompting strategies.
Instead of investigating different prompting strategies for ChatGPT, Mugdal and Wouhaybi~\cite{mudgal2023:assessment} considered seven prompts for different log-related tasks.
Among others, they considered log parsing, anomaly detection, and log prediction tasks.

From the existing studies, we can observe that the number of different LLMs investigated is low, and there is no comparison between the state-of-the-art LLMs including the free-to-use and paid proprietary models.
Moreover, there is no study exploring the analysis performance of the models pre-trained on both natural language and programming (code) language. 
We aim to fill this gap, by comparing six LLMs for the task of log parsing (see Section~\ref{subsection:llms}) including paid proprietary, free-to-use open models, and also code-specialized models.

\subsection{Evaluating Log Parsing}
\label{related:metrics}
To evaluate the performance of log parsers, several metrics have been proposed in the past.
The two most frequently applied metrics are Parsing Accuracy (PA)~\cite{jiang2024:lilac,le2023:loga,le2023:log,xu2024:divlog} and Grouping Accuracy (GA)~\cite{jiang2024:lilac,le2023:loga,le2023:log,liu2023:logprompt}.
PA, sometimes called message level accuracy, determines whether log messages have been parsed correctly~\cite{dai2020:logram}. This includes the correct replacement of variables with placeholders. In particular, the extracted template needs to agree with the ground truth.
GA, sometimes called RandIndex, determines whether log messages that correspond to the same template are parsed identically~\cite{zhu2019:tools, du2016:spell}. 
Therefore, GA is not concerned with the correct parsing of templates, but rather with a consistent parsing of log messages.

Among the two metrics, GA has received criticism~\cite{liu2022:uniparser,khan2022:guidelines,nedelkoski2021:selfsuperviseda}, leading to the proposal of other metrics for evaluating log parsers. %
Nedelkoski et al.~\cite{nedelkoski2021:selfsuperviseda} criticized grouping accuracy given that the extracted template could be completely different from the ground truth, as long as it is consistent for each log message belonging to the group.
To combat this, they proposed the use of Levenshtein Edit Distance (ED), as a means of quantifying the similarity of extracted templates and ground truth.
Edit distance has been adapted by LLM approaches~\cite{le2023:log,le2023:loga}, both of which report the median edit distance. 
Petrescu et al.~\cite{petrescu2023:log} reported the mean edit distance.

Given the relevance of string similarity metrics, such as ED, we set out to investigate the applicability of another similarity metric (i.e., longest common subsequence), which to our knowledge has not yet been investigated for log parsing.
Moreover, we provide insight into the difference between reporting mean or median values for similarity metrics, as well as on normalizing their magnitude.
Such choices can be of particular interest for the use of LLMs, which can generate diverse outputs (e.g., empty or overly long strings). %

\section{Methodology}
\begin{figure}
\includegraphics[width=\columnwidth]{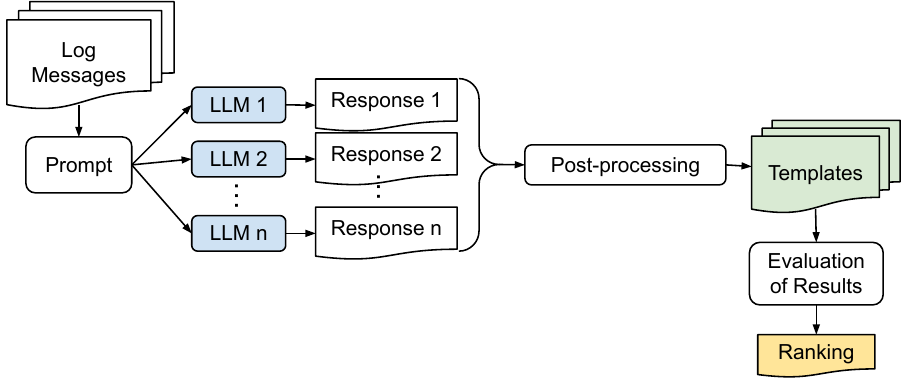}
\caption{Overview of the experimental method.}
\label{fig:overview-method}
\end{figure}

\begin{figure}
\includegraphics[width=\columnwidth]{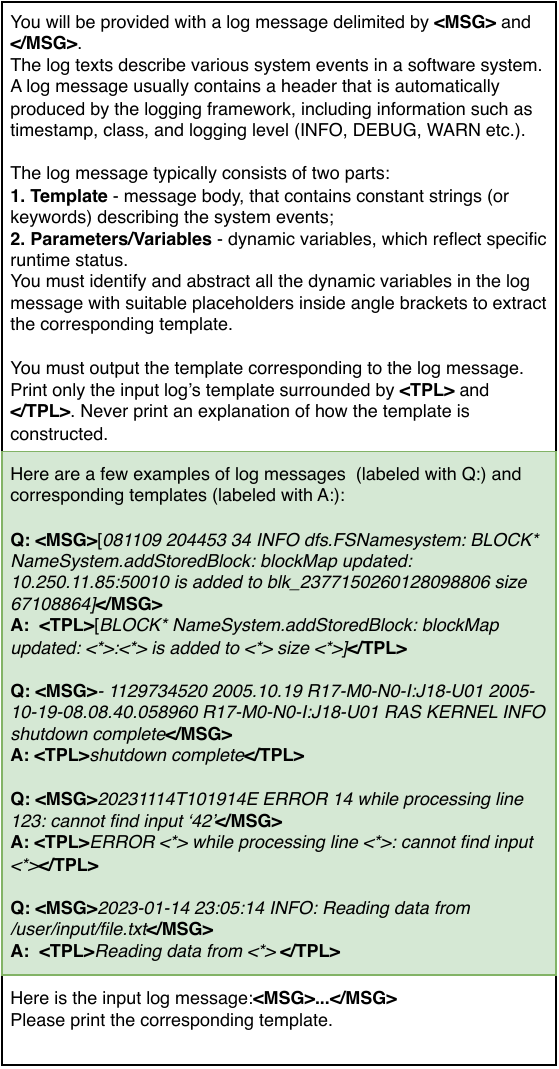}
\caption{The prompt prefix. Few-shot examples are highlighted in green. The log messages are inserted in <MSG>...</MSG> at the bottom of the prompt.}
\label{fig:prompt}
\end{figure}

Figure~\ref{fig:overview-method} gives an overview of the methodology applied in this study. 
First, we design a prompt prefix that contains a general context about log parsing and instructions to perform log parsing, in zero-shot and few-shot settings (Figure~\ref{fig:prompt}). %
We then insert a log message, for which we want to extract a template for, into the designed prompt prefix and send this prompt to a selected LLM. 
After receiving the response from the LLM, we extract the template part from the entire response. 
Subsequently, we perform post-processing on the extracted templates to normalize them with standard placeholders.
Afterward, we evaluate the results by comparing the generated templates with the ground truth in terms of their accuracy and syntactic similarity and rank the LLMs based on evaluation metrics.

\head{Prompt} 
As a first step in our study, we design prompts following approaches that have been applied in related works: zero-shot~\cite{le2023:loga, mudgal2023:assessment} and few-shot~\cite{jiang2024:lilac,le2023:log,le2023:loga,xu2024:divlog}.  
We adopt the prompt introduced by Astekin et al.~\cite{astekin2024exploratory}.
The prompts begins with a prefix, which consists of a general context about log parsing, instructions to perform the log parsing task, and an output indicator to specify the format of the extracted template.
In zero-shot prompting, we use the prompt prefix as is. 
To use few-shot prompting, we provide four examples in the prompt prefix. %
The provided examples are constructed as Q\&A pairs where each has two parts: 
the log message in the question and the corresponding template as an answer.

Figure~\ref{fig:prompt} shows both prefixes, 
where the area shaded in green is only used in the few-shot prompts, 
and the remaining area is used for both zero- and few-shot prompting.
Two of the examples are selected from the LogHub dataset~\cite{zhu2023:loghub} (Section~\ref{subsection:dataset}). However, they are different from the actual log messages used as inputs in the experiment. 
The remaining two of these four examples are taken from unrelated datasets. 
These four examples remain the same for each log message.

To identify the extracted template in the response generated by the LLM, we add instructions to the prompt to generate customized responses with specific tags. 
We specify the format of the log templates to be extracted as surrounded by a <TPL> and </TPL> pair for each corresponding input log message. 
In a similar manner, we indicate the input log message to be parsed in the form of a <MSG> and </MSG> pair.
We attach the log message to be parsed into the designed prompt prefix to construct the final form of the prompt. 
Each prompt has a single log message to be parsed as an input. 

\head{Ask LLM}
We ask LLMs to extract log templates for each of the given log messages with both zero-shot and few-shot-prompting.
To do this, we send requests to the selected set of LLMs for each prompt. 
We use the same prompt for all the LLMs and collect the generated responses. %

\head{Post-processing} 
After getting responses from the LLMs, we perform post-processing on the generated outputs to obtain the template from the responses. 
We extract the text enclosed in pairs of <TPL> and </TPL> as the raw template indicates.
In some rare cases, the responses generated by the LLMs do not include the pairs of <TPL> and </TPL>, use either <TPL> or </TPL>, 
or contain multiple <TPL> and </TPL> pairs. %
Sometimes, the LLMs return plain text without tags.
In those cases, we manually search for the raw templates in the responses and extract them if present.

We perform normalization on the raw templates before evaluating them further. 
One challenge is that LLMs at times do not follow the instructions of the defined output format, 
and use different placeholders and various formatting for the dynamic variables in the log template. 
The placeholders for the dynamic variables in the log templates often have different names and formats. 
Even if we define the format to show the dynamic variables inside angle brackets, 
we encounter various formats (curly brackets, square brackets, dollar signs, etc.). 
To address this problem, we investigate all the raw templates to detect formatting issues and construct regular expressions to automatically convert them into the expected format. 

\head{Evaluation of Results} 
After obtaining normalized templates, we measure the quality of the extracted templates in two main aspects: \emph{accuracy} and \emph{syntactic similarity}.
The \emph{accuracy} is evaluated with the number of correctly parsed log messages, where a log message is considered correctly parsed only when each token in the log message is correctly recognized as a template or variable based on the ground truth.
The \emph{syntactic similarity} is measured by comparing the text similarity between the generated templates and ground truth.
It is determined based on Edit Distance, Longest Common Subsequence, and their variants. 
More details about the metrics are provided in Section~\ref{subsection:evaluation_metrics}.
We provide the ranking of LLMs in terms of each evaluation metric.

\section{Experimental Design}
\subsection{Research Questions}
We address the following research questions in our empirical study:

\textbf{RQ1. How do the different LLMs perform for log parsing?}

\noindent
In the first research question, we set out to assess the performance of six LLMs (Section~\ref{subsection:llms}) based on three metrics (Section~\ref{subsection:evaluation_metrics}) on a real-world dataset (Section~\ref{subsection:dataset}). 
We provide details on the performance of LLMs for each of the three metrics and determine which models perform best.
Moreover, we compare the performance of paid and free-to-use models. 

\textbf{RQ2: What is the impact of the studied metrics on the ranking of LLMs?}

\noindent
After determining the best-performing LLMs according to three metrics in RQ1, we investigate how influenceable the ranking of LLMs by the investigated metrics is.
For instance, the reported similarity metrics in RQ1 are reported based on median values, and RQ2 checks whether the ranking of LLMs changes when reporting mean values (see Section~\ref{subsection:evaluation_metrics} for more details).

\textbf{RQ3: How easy to use are the generated responses by the LLMs?}

\noindent
While RQ1 and RQ2 are concerned with the performance of LLMs, RQ3 is concerned with the usability. 
In particular, we are interested in how easy it is to use the generated outputs of LLMs to extract the intended part from the entire output.
Therefore, we provided qualitative insights into the manual work involved in evaluating the outputs generated by each of the six LLMs.

\begin{table}
\caption{Description of the LogHub dataset. Number of unique templates for the original and corrected datasets (Khan et al.~\cite{khan2022:guidelines}) are shown.}
\label{table:dataset}
\centering
\scalebox{0.8}{
\begin{tabular}{llra}
\toprule
\rowcolor{white}
Type & Project & \# Original~~\cite{zhu2023:loghub} & \# Khan et al.~\cite{khan2022:guidelines} \\ \midrule
\multirow{5}{*}{Distributed Systems} & HDFS & 14 & 14 \\
 & Hadoop & 114 & 114 \\
 & Spark & 36 & 36 \\
 & Zookeeper & 50 & 50 \\ 
 & OpenStack & 43 & 43 \\ \midrule
\multirow{3}{*}{Super computers} & BGL & 120 & 120 \\
 & HPC & 46 & 46 \\
 & Thunderbird & 149 & 149 \\ \midrule
\multirow{3}{*}{Operating systems} & Windows & 50 & 50 \\
 & Linux & 118 & 118 \\
 & Mac & 341 & 341 \\ \midrule
\multirow{2}{*}{Mobile systems} & Android & 166 & 158 \\
 & HealthApp & 75 & 75 \\ \midrule
\multirow{2}{*}{Server applications} & Apache & 6 & 6 \\
 & OpenSSH & 27 & 26 \\ \midrule
Standalone software & Proxifier & 8 & 8 \\ \midrule
Total &  & 1363 & 1354 \\ \bottomrule
\end{tabular}
}
\end{table}

\subsection{Dataset}
\label{subsection:dataset}
To evaluate the log parsing performance of LLMs, we apply them to LogHub~\cite{zhu2023:loghub}, a dataset of real-world log messages.
These messages originate from different systems (i.e., distributed systems, supercomputers, operating systems, mobile systems, server applications, standalone software) distributed over 16 projects.
For each of the 16 projects, LogHub provides 2,000 log messages each linked to a corresponding log template that should be extracted. 
In this context, multiple log messages of a project can have the same template.

An overview of the 16 projects and the unique number of log templates is shown in Table~\ref{table:dataset}.
In addition to listing the original number of unique log templates, we provide the number of templates from a modified LogHub version by Khan et al.~\cite{khan2022:guidelines}.
Like related works~\cite{xu2024:divlog,le2023:log,le2023:loga}, we consider the modified templates by Khan et al.~\cite{khan2022:guidelines}.
Among others, they replaced hard-coded boolean values ($true$, $false$) with placeholders. This leads to a reduction of the total number of templates in the dataset from 1,363 to 1,354.
Therefore, we carry out experiments with 1,354 log messages, one message for each of the templates.
This allows us to investigate LLMs on a large variety of log templates while reducing the total number of log messages evaluated and keeping manual effort for the post-processing within reason.

\subsection{Large Language Models}
\label{subsection:llms}
Table~\ref{tab:llms} presents an overview of the LLMs used in this study. 
The columns indicate the version, availability, and base model, in case it is built by fine-tuning another LLM.
Since our goal is to explore the capability of state-of-the-art LLMs in the log parsing task, we select LLMs with different availability, parameter size, and fine-tuning options.
The selected LLMs contain both closed-source and publicly accessible models, including base and fine-tuned models with different parameter sizes. 

\begin{table}
\centering
\caption{Overview of large language models.}
\label{tab:llms}
\scalebox{0.85}{
\rowcolors{1}{}{lightgray}
\begin{tabular}{llllll}
\toprule
Model       & \begin{tabular}[c]{@{}l@{}}Version/\\Release time\end{tabular}               & Base model  & Availability        \\ \toprule
GPT-3.5     & \begin{tabular}[c]{@{}l@{}}gpt-3.5-turbo-0613/\\June, 2023\end{tabular}      & -                & Closed-source       \\ %
Claude 2.1    & \begin{tabular}[c]{@{}l@{}}November, 2023\end{tabular}                   & -               & Closed-source       \\ %
Llama 2    & \begin{tabular}[c]{@{}l@{}}llama2:7b/\\July, 2023\end{tabular}               & -             & Public   \\ %
CodeLlama   & \begin{tabular}[c]{@{}l@{}}codellama:7b-instruct/\\August, 2023\end{tabular} & Llama 2       & Public   \\ %
Zephyr Beta & \begin{tabular}[c]{@{}l@{}}zephyr:7b/\\October, 2023\end{tabular}            & Mistral-7B       & Public   \\ %
CodeUp      & \begin{tabular}[c]{@{}l@{}}codeup:13b/\\July, 2023\end{tabular}              & Llama 2           & Public  \\ \bottomrule
\end{tabular}
}
\end{table}

Of the six models, two (GPT-3.5\footnote{~\url{https://platform.openai.com/docs/models/gpt-3-5}} and Claude 2.1\footnote{~\url{https://www.anthropic.com/news/claude-2-1}}) are proprietary, paid-for models that can be accessed via API calls. 
The other four LLMs (Llama2~\cite{touvron2023:llama}, CodeLlama~\cite{roziere2023:code}, Zephyr~\cite{tunstall2023:zephyr}, CodeUp~\cite{jiang2023:codeup}) are publicly accessible and available for download from Hugging Face.\footnote{~\url{https://huggingface.co/}}
In our implementation, we utilized Ollama\footnote{~\url{https://ollama.ai/}} to access the free-to-use models and used the APIs provided to access the paid proprietary models.\footnote{~The costs for running these experiments were low: the GPT-3.5 experiments cost approx. USD 3, and the costs for Claude 2.1 were USD 1.19.}

Among the six models, GPT-3.5, Claude 2.1, and Llama2 are base models, while Zephyr, CodeUp, and CodeLlama are fine-tuned models.
Zephyr is fine-tuned for improved chat and is stated as the best open-access RLHF-based model~\cite{tunstall2023:zephyr}.
CodeUp and CodeLlama are fine-tuned for code generation tasks. 
Paid proprietary models contain a large number of parameters, where Claude 2.1 contains approximately 200 billion and GPT-3.5 has 175 billion parameters.
The variants of Llama2, CodeLlama, and Zephyr are used in sizes of 7 billion parameters, and CodeUp with 13 billion parameters. 

Lower temperatures make the LLMs more targeted and deterministic.
Considering the trade-off between coherence and creativity for the log parsing task, we set their temperatures to 0.2 so that LLMs would generate more deterministic outputs for the same type of prompts.

\subsection{Evaluation Metrics}
\label{subsection:evaluation_metrics}
In accordance with related work, we use PA to evaluate the accuracy of LLMs for log parsing~\cite{jiang2024:lilac,le2023:loga,le2023:log,xu2024:divlog}.
Given the circumstance that we consider a single log message per template, we are not able to measure grouping accuracy.
To measure the textual similarity of extracted templates $T$ and the ground truth $GT$, we use ED and LCS.

In addition to reporting ED and LCS, we also consider their normalized variants.
For this purpose, we compute Edit Similarity (ES) instead of Edit Distance, such that similarity is maximized at 1 and minimized at 0, in accordance with normalized LCS and PA:
\begin{equation}
    ES_{norm} = 1 - (ED(T, GT) / max(len(T), len(GT)))
\end{equation}
\begin{equation}
    LCS_{norm} = LCS(T,GT) / len(GT)
\end{equation}

For RQ1, we report the median values for ED and LCS, in accordance with related works~\cite{le2023:log,le2023:loga}.
For RQ2, we consider normalized and standard ED and LCS, as well as the median and mean values.
To better compare normalized ED and LCS with PA, we report their sum rather than mean.
In total, we consider four variants for each similarity metric:
standard-median, standard-mean, normalized-median, normalized-sum.

\subsection{Threats to Validity}
There are three factors impacting the threats to validity of our study: LLMs, dataset, and metrics.

We considered a total of six LLMs for our study, which extends the number of LLMs investigated in previous works.
Four out of six LLMs are publicly accessible. We report the version we used to allow for reproducible results.
Closed source models may not be as reproducible given their nature (i.e., GPT-3.5 and Claude 2).
For generating outputs, we employed a temperature of $0.2$. 
We note that this value is chosen arbitrarily, as we observed varying temperatures applied in research articles and Hugging Face implementations. %
To limit manual work, we focused on a single temperature but note that a deeper investigation of other values could be beneficial.

We measured a total of three metrics, two of which are applied in related works (ED, PA) and have shown useful in assessing log parsers.
For LCS, which to the best of our knowledge is used to evaluate log parsing for the first time, we compare it to ED and provide insight on when the use of LCS is preferable (RQ2). This supports the applicability of LCS to log parsing.

The dataset we used is a subset of the popular LogHub dataset. 
Given the manual post-processing required when using LLMs (RQ3), we chose a single log message per available template to test LLMs on the largest variety of log templates, while keeping manual post-processing within reason.

\section{Results and Discussion}

\begin{table*}
\caption{Parsing accuracy comparison. The best values are highlighted in bold. For each model, we provide results for zero-shot (Zero) and few-shot (Few) prompts. }
\label{table:parsing}
\centering
\scalebox{0.8}{
\begin{tabular}{lbbaabbaabbaa}
\toprule
 & \multicolumn{2}{c}{GPT-3.5} & \multicolumn{2}{c}{Claude 2.1} & \multicolumn{2}{c}{Llama 2} & \multicolumn{2}{c}{CodeLlama} & \multicolumn{2}{c}{CodeUp} & \multicolumn{2}{c}{Zephyr} \\
 \rowcolor{white}
 & Zero & Few & Zero & Few & Zero & Few & Zero & Few & Zero & Few & Zero & Few \\ \midrule
HDFS & 0 & 2 & 3 & \textbf{5} & 2 & 3 & 2 & 2 & 2 & 3 & 1 & 1 \\
Hadoop & 16 & 39 & 30 & \textbf{50} & 15 & 10 & \textbf{50} & 43 & 15 & 22 & 8 & 6 \\
Spark & 10 & \textbf{21} & 13 & 20 & 11 & 11 & 18 & 17 & 9 & 20 & 7 & 2 \\
Zookeeper & 8 & 27 & 19 & 25 & 9 & 11 & \textbf{31} & 29 & 8 & 22 & 3 & 4 \\
OpenStack & 8 & 13 & 6 & 25 & 6 & 9 & \textbf{28} & 25 & 5 & 3 & 2 & 2 \\ %
BGL & 11 & 41 & 14 & \textbf{66} & 1 & 4 & 8 & 32 & 0 & 8 & 0 & 6 \\
HPC & 0 & 3 & 0 & 1 & 0 & 7 & 7 & \textbf{10} & 1 & 4 & 2 & 1 \\
Thunderbird & 18 & 45 & 13 & 21 & 0 & 30 & 19 & \textbf{50} & 19 & 37 & 0 & 0 \\ %
Windows & 17 & \textbf{21} & 15 & 15 & 1 & 13 & 13 & 19 & 0 & 3 & 12 & 0 \\
Linux & 41 & 64 & 29 & 48 & 26 & 41 & 60 & \textbf{76} & 41 & 66 & 20 & 12 \\
Mac & 96 & 125 & 105 & 55 & 29 & 53 & 106 & \textbf{130} & 62 & 105 & 64 & 25 \\ %
Android & 47 & 77 & 72 & 81 & 6 & 56 & 73 & \textbf{91} & 22 & 58 & 5 & 3 \\
HealthApp & 28 & 33 & 21 & 19 & 0 & 25 & 39 & \textbf{47} & 6 & 34 & 2 & 1 \\ %
Apache & 1 & \textbf{5} & 0 & 0 & 0 & 0 & 4 & 3 & 1 & 0 & 0 & 0 \\
OpenSSH & 8 & \textbf{15} & 6 & 8 & 1 & 8 & 12 & 12 & 2 & 8 & 0 & 0 \\ %
Proxifier & 0 & 0 & 0 & 0 & 0 & 0 & 0 & 0 & 0 & 0 & 0 & 0 \\ \midrule
Total & 309 & 531 & 346 & 439 & 107 & 281 & 470 & \textbf{586} & 193 & 393 & 126 & 63  \\ \bottomrule
\end{tabular}
}
\end{table*}

\begin{table*}
\caption{Edit distance and longest common subsequence comparison. This table shows median values, with the best scores highlighted in bold. Edit distance is to be minimized; longest common subsequence is to be maximized. For each model, we provide results for zero-shot (Zero) and few-shot (Few) prompts. }
\label{table:similarity}
\centering
\scalebox{0.72}{
\begin{tabular}{lbbaabbaabbaa|b|rbaabbaabbaa}
\toprule
 & \multicolumn{12}{c|}{Edit Distance - Median (small is better)} &  & \multicolumn{12}{c}{Longest Common Subsequence - Median (large is better)} \\
 & \multicolumn{2}{c}{GPT-3.5} & \multicolumn{2}{c}{Claude 2.1} & \multicolumn{2}{c}{Llama 2} & \multicolumn{2}{c}{CodeLlama} & \multicolumn{2}{c}{CodeUp} & \multicolumn{2}{c|}{Zephyr} &  & \multicolumn{2}{c}{GPT-3.5} & \multicolumn{2}{c}{Claude 2.1} & \multicolumn{2}{c}{Llama 2} & \multicolumn{2}{c}{CodeLlama} & \multicolumn{2}{c}{CodeUp} & \multicolumn{2}{c}{Zephyr} \\ 
 \rowcolor{white}
 & Zero & Few & Zero & Few & Zero & Few & Zero & Few & Zero & Few & Zero & Few &  & Zero & Few & Zero & Few & Zero & Few & Zero & Few & Zero & Few & Zero & Few \\ \midrule
HDFS & 13.0 & 11.0 & 8.0 & \textbf{3.0} & 23.0 & 8.5 & 4.0 & 8.0 & 18.0 & 6.0 & 29.0 & 15.5 &  & 38.5 & 37.5 & 38.5 & \textbf{41.5} & 25.0 & 39.5 & 41.0 & 39.5 & 36.5 & 41.0 & 30.5 & 33.0 \\
Hadoop & 22.0 & 8.0 & 8.0 & 4.0 & 26.5 & 13.0 & \textbf{3.5} & \textbf{3.5} & 70.5 & 15.0 & 55.0 & 59.5 &  & \textbf{41.0} & 37.0 & 36.0 & 36.0 & 22.0 & 34.0 & 34.0 & 34.0 & 35.0 & 36.5 & 31.5 & 32.0 \\
Spark & 15.5 & \textbf{0.0} & 6.5 & \textbf{0.0} & 20.0 & 5.5 & 1.5 & 2.5 & 44.5 & \textbf{0.0} & 26.5 & 28.0 &  & 37.0 & 37.0 & 32.0 & 30.5 & 19.5 & 29.5 & 34.5 & 35.0 & 34.0 & \textbf{38.5} & 35.0 & 9.5 \\
Zookeeper & 3.5 & \textbf{0.0} & 3.5 & \textbf{0.0} & 29.5 & 9.0 & \textbf{0.0} & \textbf{0.0} & 89.0 & 3.0 & 51.5 & 38.5 &  & 35.5 & \textbf{38.0} & \textbf{38.0} & 36.0 & 10.5 & 32.5 & 37.0 & \textbf{38.0} & 31.0 & 37.0 & 28.0 & 30.5 \\
OpenStack & 16.0 & 16.0 & 85.0 & \textbf{0.0} & 36.0 & 18.0 & \textbf{0.0} & \textbf{0.0} & 34.0 & 18.0 & 71.0 & 43.0 &  & 35.0 & 36.0 & 46.0 & \textbf{48.0} & 13.0 & 39.0 & 46.0 & 46.0 & 38.0 & 42.0 & 36.0 & 27.0 \\
BGL & 20.0 & 3.0 & 63.0 & \textbf{0.0} & 44.0 & 13.0 & 20.0 & 3.0 & 51.5 & 12.0 & 49.5 & 44.5 &  & 44.0 & 44.0 & 44.0 & \textbf{45.0} & 6.0 & 44.0 & 44.0 & 44.0 & 41.0 & 43.5 & 40.0 & 41.0 \\
HPC & 33.5 & 24.5 & 38.0 & 29.0 & 31.0 & \textbf{18.0} & 22.0 & 22.5 & 35.0 & 25.0 & 37.5 & 34.0 &  & 22.5 & \textbf{24.0} & 22.0 & 23.0 & 6.0 & 18.5 & 17.5 & 22.5 & 10.0 & 18.0 & 18.0 & 7.5 \\
Thunderbird & 10.0 & 6.0 & 13.0 & 12.0 & 26.0 & 8.0 & 9.0 & \textbf{5.0} & 19.0 & 10.0 & 25.0 & 33.0 &  & 28.0 & \textbf{29.0} & \textbf{29.0} & 27.0 & 6.0 & 28.0 & 28.0 & \textbf{29.0} & 27.0 & 27.0 & 13.0 & 10.0 \\
Windows & 10.0 & \textbf{4.5} & 6.5 & 7.5 & 37.5 & 10.0 & 7.0 & 5.0 & 50.0 & 13.5 & 28.0 & 36.5 &  & \textbf{45.0} & \textbf{45.0} & 40.5 & \textbf{45.0} & 14.5 & 41.0 & 41.5 & 43.5 & 39.0 & 39.5 & 42.5 & 37.5 \\
Linux & 3.5 & \textbf{0.0} & 9.0 & 4.5 & 17.0 & 5.0 & \textbf{0.0} & \textbf{0.0} & 5.5 & \textbf{0.0} & 30.0 & 30.0 &  & 31.5 & 31.5 & 28.0 & 30.0 & 18.5 & 27.0 & 28.5 & 30.0 & 29.0 & \textbf{32.0} & 31.0 & 25.0 \\
Mac & 10.0 & \textbf{6.0} & 14.0 & 22.0 & 38.0 & 24.0 & 14.0 & 7.0 & 32.0 & 10.0 & 40.0 & 43.0 &  & 50.0 & 49.0 & \textbf{56.0} & 47.0 & 15.0 & 37.0 & 47.0 & 48.0 & 47.0 & 49.0 & 44.0 & 34.0 \\
Android & 6.0 & \textbf{0.0} & 3.0 & \textbf{0.0} & 25.0 & 4.0 & 2.0 & \textbf{0.0} & 17.0 & 4.0 & 25.0 & 34.0 &  & 38.0 & 38.5 & 37.5 & 37.5 & 17.0 & 36.5 & 38.5 & \textbf{40.5} & 34.0 & 37.0 & 31.0 & 19.5 \\
HealthApp & 3.0 & 2.0 & 20.0 & 18.0 & 35.0 & 5.0 & \textbf{0.0} & \textbf{0.0} & 49.0 & 1.0 & 49.0 & 42.0 &  & \textbf{35.0} & \textbf{35.0} & 32.0 & 35.0 & 3.0 & 31.0 & 34.0 & 31.0 & 30.0 & \textbf{35.0} & 24.0 & 15.0 \\
Apache & 8.5 & \textbf{0.0} & 14.5 & 12.0 & 33.0 & 28.0 & \textbf{0.0} & 4.0 & 31.0 & 8.5 & 17.5 & 35.0 &  & 36.5 & 39.5 & 39.0 & 35.0 & 6.0 & 28.5 & 39.5 & 40.0 & 38.0 & 40.0 & \textbf{43.0} & 33.0 \\
OpenSSH & 7.0 & \textbf{0.0} & 14.0 & 11.0 & 41.0 & 15.5 & 6.0 & 3.0 & 20.5 & 7.0 & 42.0 & 45.0 &  & 46.0 & \textbf{50.5} & 46.0 & 49.0 & 9.0 & 42.0 & 44.5 & 50.0 & 40.0 & 44.5 & 43.5 & 37.5 \\
Proxifier & 48.5 & 23.0 & 17.0 & 25.5 & 23.5 & 20.5 & 18.0 & \textbf{16.0} & 52.0 & 54.0 & 45.5 & 69.5 &  & 64.0 & 66.5 & 58.0 & \textbf{76.0} & 44.0 & 45.5 & 54.0 & 61.0 & 54.5 & 61.0 & 33.5 & 18.5 \\ \midrule
Total & 11.0 & 4.0 & 14.0 & 11.0 & 31.0 & 11.0 & 7.0 & \textbf{3.0} & 36.0 & 9.0 & 35.0 & 37.0 &  & 38.0 & \textbf{39.0} & \textbf{39.0} & 38.0 & 12.0 & 35.0 & 37.0 & 38.0 & 35.0 & 38.0 & 33.0 & 28.0 \\
\bottomrule
\end{tabular}
}
\end{table*}
\subsection{RQ1: Performance Comparison}

\head{Parsing accuracy}
Table~\ref{table:parsing} compares the PA of the six LLMs.
For this purpose, we sum up the number of correctly parsed templates for each of the 16 projects.
We can observe that CodeLlama is the best performing LLM for log parsing based on PA, with 586 out 1,354 log messages parsed correctly when using the few-shot prompt.
The best performing LLM is CodeLlama for 9 out of 16 projects.
The next best-performing LLMs are GPT3.5 and Claude 2, the two paid models, both under consideration of few-shot prompts. 

Generally speaking, the results indicate that few-shot prompts allow for a better performance of LLMs in 5 out of 6 cases.
Only for Zephyr, the zero-shot prompt is able to achieve a better performance than the use of few-shot prompts.
One reason might be that Zephyr is not able to use the additional information but rather gets confused by the long text. 
Supporting this assumption, there are 56 cases where Zephyr-few does not return an output at all, and 364 cases where it returns the prompt template.

Moreover, none of the LLMs are able to correctly parse a log message for the Proxifier project. 
This could be explained by the low number of messages, given the 8 distinct log templates.
However, this shows that PA does not always allow for a comparison between methods and highlights the importance of other metrics, such as ED and LCS.
With these two metrics, one can assess how good the extracted templates are, even when none of them is 100\% correct.

\begin{myframed}
\noindent
CodeLlama is the best model based on Parsing Accuracy.
\end{myframed}

\head{Edit distance and Longest common subsequence}
Table~\ref{table:similarity} presents the ED and LCS results for the six LLMs.
We report median values for each of the 16 projects, as well as the median overall 1,354 in the ``Total'' column. 
Here, ED should be minimized, whereas LCS should be maximized to indicate good performance.

The left half of Table~\ref{table:similarity} shows results for ED. 
Similar to PA, the best performing LLM is CodeLlama, with a median ED of $3$ and the best results on 9 out of 16 projects.
CodeLlama is followed by GPT3.5-few, which has a median ED of $4$.
Unlike the results for PA, the third ranking LLM is CodeUp-few, performing better than Claude 2.

We can also observe that the use of median ED values leads to ties among the LLMs, in particular when an LLM is able to extract more than 50\% of the templates correctly, leading to an ED of $0$.

\begin{myframed}
\noindent
CodeLlama is the best performing model based on Edit Distance.
\end{myframed}

The right half of Table~\ref{table:similarity} presents the median LCS for the six LLMs over the 16 projects. 
We observe a tie for the best performing model, with GPT3.5-few and Claude 2-zero exhibiting a median LCS of $39$.
These LLMs are followed by CodeLlama-few, CodeUp-few, GPT3.5-zero and Claude 2-few, all of which share an LCS of $38$.
The lowest LCS is achieved by Llama 2 and Zephyr.

Similar to the reported median values for ED, we observe ties among the LLMs for LCS as well. In this case, ties happen for the best performing models, which makes it harder to distinguish LLMs based on their performance.

\begin{myframed}
\noindent
Paid models (GPT-3.5, Claude 2.1) are the best performing models based on Longest Common Subsequence.
\end{myframed}

\head{Paid vs. free-to-use LLMs}
CodeLlama, a free model,, is the best performing on metrics employed by existing work: PA, ED.
Paid models, in particular GPT-3.5, are only able to outperform free-to-use models when measuring LCS, to our knowledge first used for log parsing in this publication.
CodeLlama shares the second rank for LCS and would therefore be a feasible option as well.
In particular, free-to-use LLMs trained on code have achieved good results (CodeUp and CodeLlama), outperforming Llama 2 and Zephyr. %

This shows that one does not need to spend money on paid LLMs to research log parsing, as free models, such as CodeLlama can even outperform them for two out of three metrics in our experiments. The only downside of using free models is that one needs to be able to load them (e.g., CodeLlama has 7 billion parameters) and have a modern processing unit, such as a GPU, to obtain results quicker. 
If hardware is a limiting factor, GPT3.5 and Claude 2.1 are promising alternatives. 
\begin{myframed}
\noindent
Free models, such as CodeLlama, are a high-performing alternative to paid APIs for log parsing.
\end{myframed}

\begin{figure}[t]
\includegraphics[width=\columnwidth]{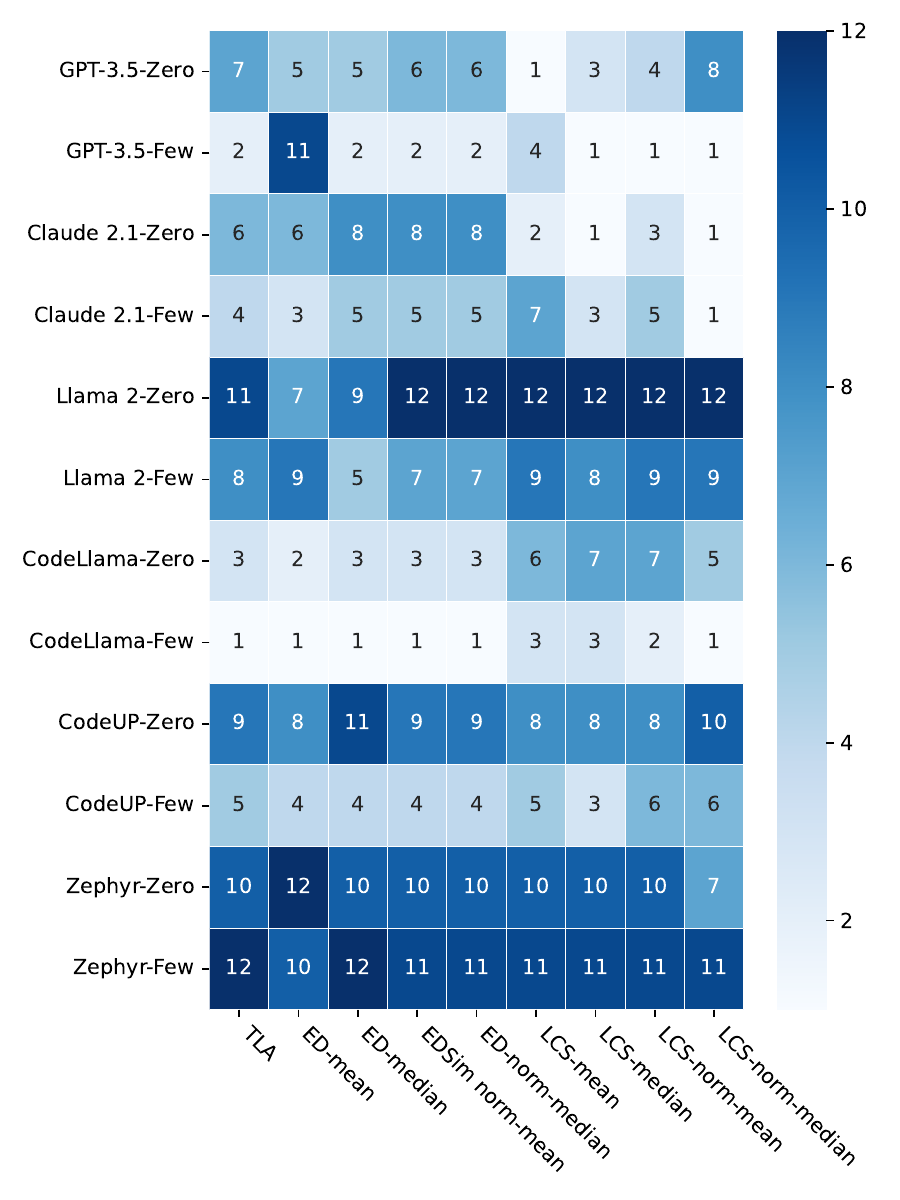}
\caption{Ranking based on metrics.}
\label{fig:ranking}
\end{figure}

\begin{figure*}
\centering
\begin{minipage}{\linewidth}
 \subfloat[Mean and median. \\Correlation coefficient = 0.83]{%
  \includegraphics[width=0.32\linewidth]{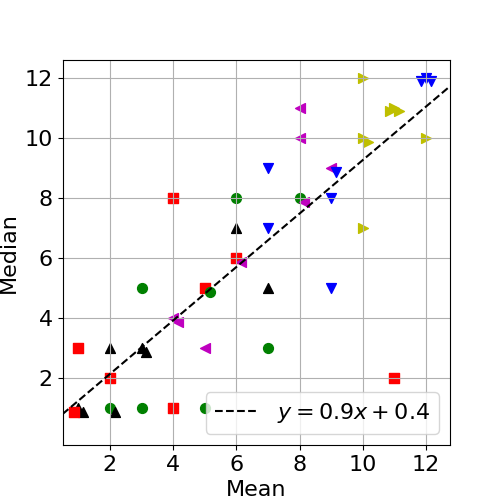}
}
\subfloat[Standard and normalized. \\Correlation coefficient = 0.72]{%
 \includegraphics[width=0.32\linewidth]{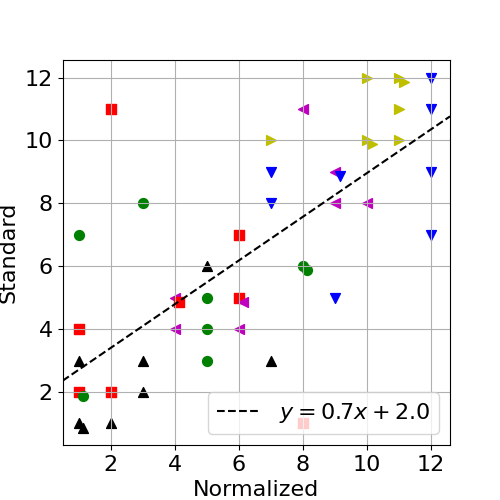}
 }
  \subfloat[Edit distance and Longest Common Subsequence. Correlation coefficient = 0.68]{%
 \includegraphics[width=0.32\linewidth]{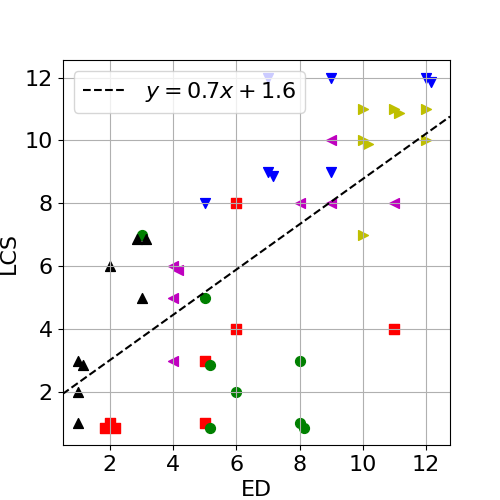}
 }
\end{minipage}

\begin{minipage}{\linewidth}
\centering
  \includegraphics[width=0.3\linewidth]{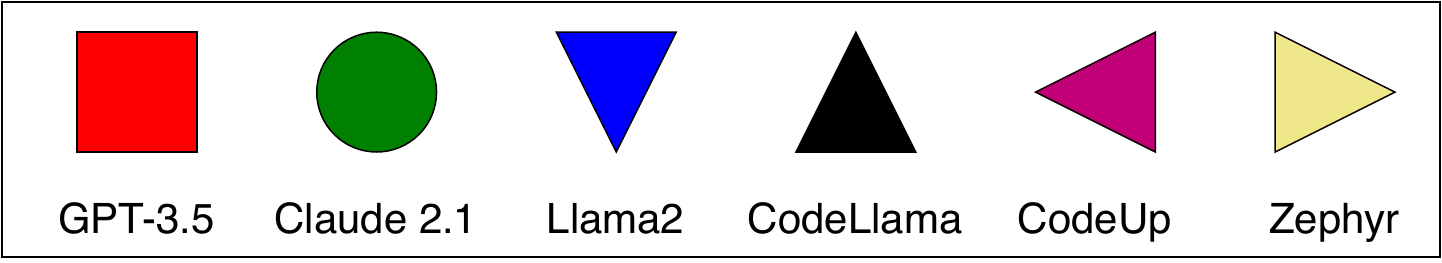}
\end{minipage}

\caption{Comparison of different aspects based on ranking with Pearson correlation coefficient. Colour and shape show the different LLMs.}
\label{fig:comparison}
\end{figure*}

\subsection{RQ2: Impact of Metrics}

\head{Ranking Stability}
First, we investigate the ranking of LLMs according to their performance over different metrics and whether their rank changes.
In total, we report nine metrics: PA, four variants of ED, and four variants of LCS.
The four variants of ED and LCS are median, mean, normalized median, and normalized mean (Section~\ref{subsection:evaluation_metrics}).
This allows us to investigate ranking differences between metrics (PA, ED, LCS) and differences based on reporting choices (mean, median, normalizing), all of which can impact which LLM is perceived as best performing.

The ranking of each of the 6 LLMs, given two prompting types (i.e., zero- and few-shot), for each of the nine metrics is shown in Figure~\ref{fig:ranking}. Here, rank $1$ signifies the best model, while rank $12$ is the worst-performing one.
We observe that the results for best and worst performing LLMs are consistent over similarity metrics. 
The best performing LLM for all four ED variants is CodeLlama-few. 
GPT3.5 is the highest ranking model for LCS; in two out of four cases there is a tie for first place with Claude 2.1.
The worst model for each metric is either Zephyr or Llama 2.

The ranking over different metrics can be stable with some models deviating by 2-3 ranks (i.e., Zephyr-few, CodeUp-few, CodeUp-zero, CodeLlama-few) over all 9 computed metrics.
However, we can also notice high deviations, in particular ones caused by computing mean values.
For example, GPT3.5-few, which ranks among the 1st-4th best models for 8 out of 9 metrics, is ranked 11th when computing mean edit distances. 
GPT3.5-zero which was 3rd ranked at best for 8 metrics, is the best on LCS-mean.
Such deviations when computing means can be caused by outliers.
For GPT3.5-few, this happens on the Hadoop project, where the following template is extracted:
\begin{lstlisting}[basicstyle= \small,frame = single]
attempt_<*><*><*><*><*><*><*><*> .... <*><*>     
\end{lstlisting}
<$*$> is repeated multiple times such that the edit distance for a single template went up to $5,272$.
Reporting normalized mean values for ED and LCS does not suffer from such outliers, as values are in a range of [0,1].

\begin{myframed}
\noindent
Do not report mean values for LLMs without normalizing. 
\end{myframed}

\head{Metric comparison}
Next, we take a closer look at what choices can impact ranking deviations. In particular, we consider three comparisons for similarity metrics: ED and LCS, mean and median, standard and normalized. 
The purpose of these comparisons is to determine whether and how much of a difference the corresponding choices have on the results (i.e., overall rank) of the LLMs. 

Figure~\ref{fig:comparison} visualizes the three comparisons.
For each comparison, we compare the ranks of LLMs based on the two options, each determined by one of the axes in the plots.
For each model, we visualize 8 markers (2*4), for the two prompt types and four metric variants of ED and LCS.
Moreover, we report the Pearson correlation coefficient between ranks over all LLMs.
If we find a strongly correlated pair, the choice between both alternatives (e.g., mean or median) would not be critical for the results. 
If however there is no correlation between the two options, then one needs to be careful when picking metrics for evaluation.

Median and mean have the highest correlation even though the mean is impacted by outliers, such as the ones observed for GPT3.5 (i.e., a single log message with ED of $5,272$).
For traditional approaches, e.g. regex over log messages, it might not make a difference which to use, but due to the stochastic nature of LLMs it is preferred to use median over mean to avoid outliers (or normalized mean for this purpose).
One disadvantage that can be avoided by using mean values is the lower chance of ties when ranking LLMs, in particular when more than 50\% of the templates are extracted correctly.
For traditional approaches, Petrescu et al.~\cite{petrescu2023:log} reported mean ED values. %

\begin{myframed}
\noindent
Median loses meaning when more than 50\% of the log messages are solved correctly. Mean values take every log message into account.
\end{myframed}

The ranks of standard and normalized values have a Pearson correlation coefficient of $0.72$.
Similar to the comparison of mean and median, this can be impacted by outliers, with normalized values being bound within a range of [0,1].
By normalizing LCS and ED, each log message has the same level of impact, independent of length.
This is similar to PA, where a correctly extracted template counts as 1 while an incorrect template as 0.
Using standard ED measures, an empty message corresponding to a long ground truth template would receive a worse score than a short one, even though they are incorrect to the same degree.

\begin{lstlisting}[basicstyle= \small,frame = single]
$example_1$ = ""
$template_1$ = queue: default
ED = 14
$ED_{norm}$ = 1
    
$example_2$ = ""
$template_2$ = ciod: Error creating node map 
            from file <*>: No child processes
ED = 63
$ED_{norm}$ = 1
\end{lstlisting}

\begin{myframed}
\noindent
Normalizing avoids outliers and gives every log message the same importance.
\end{myframed}

ED and LCS are the least correlated pair in our comparison, which can be explained by differences in dealing with timestamp/message scope as well as placeholders:
\begin{lstlisting}[basicstyle= \small,frame = single]
$example_1$ = CBS    SQM: Cleaning up report files 
                            older than <*> days.
$template_1$ = SQM: Cleaning up report files 
                            older than <*> days.
$ES_{norm}$ = 0.88
$LCS_{norm}$ = 1
    
$example_2$= setDataSource(<*>, <*>, <*>)
$template_2$= setDataSource(166, 0, 576460752303423487)
$ES_{norm}$ = 0.41
$LCS_{norm}$ = 0.68
\end{lstlisting}

From this example, we can see that LCS ignores timestamps (excess tokens as long as the ground truth is part of the message) while ED would remove these. Note that we report ES to better compare with LCS, as both are maximized at $1$.
When an LLM fails to replace a variable by the placeholder <$*$>, LCS is not impacted by the length of the variable, ED on the other hand is. 
The use of LCS and ED should therefore be made based on the application scenario, or both can be considered.

\begin{myframed}
\noindent
Edit distance is preferred when timestamps should be removed (scope). Longest common subsequence is preferred when the detection of placeholders is of higher importance.
\end{myframed}

\subsection{RQ3: Usability of Responses}
We investigate how usable the LLM responses are when extracting log templates.  %
The usability of a response is a qualitative indication of the clarity to distinguish the log template in the correct format from the generated response.
Even though we explicitly describe the expected response format in the prompt, there are still many responses that do not meet the specifications (e.g., responses with <TPL> and </TPL>). %

It is a challenging task to find the placeholders in the generated log templates since LLMs produce them in different formats and names than defined in the prompt.
In the zero-shot prompt, we do not specify the placeholder for the dynamic variables, we only instruct to print the dynamic variables with suitable placeholders inside angle brackets. 
In this way, we aim to observe their preference for finding suitable placeholder names. 
However, we observe that none of the LLMs are able to follow the formatting of angle brackets <> consistently in zero-shot prompting. 
GPT-3.5 zero-shot experiment produces the best results, following the formatting of angle brackets <> with only 7 exceptions of log messages. 
Other models use a variety of brackets such as (double) curly brackets, (double) square brackets, tags enclosed in left and right angle brackets, @ sign, dollar sign, etc.
More details on the usability of responses with zero-shot prompting are provided in Section~\ref{subsection:observations_zero-shot}.

In the few-shot experiments, we give four examples of the expected output (<$*$>) while we still instruct LLMs to print the variables inside angle brackets.
With these examples, GPT-3.5, Claude 2.1, CodeLlama, and CodeUp are able to format variables using <$*$> with only a few exceptions.
More details about the few-shot experiments of the LLMs are provided in Section~\ref{subsection:observations_few-shot}.

We present our observations from both zero-shot and few-shot experiments in terms of two aspects: LLMs' placeholder selection and variable formatting. 

\subsubsection{Observations from zero-shot experiments} 
\label{subsection:observations_zero-shot}
\noindent

\head{Placeholder selection}
GPT-3.5 and Claude 2.1 predominantly employ generic placeholder names like VAR, variable1, and variable2.
Frequently, they opt for the repetition of preceding identifier names as placeholders, as exemplified by constructs such as lock = <lock> and flag = <flag>.
Besides that, they are able to generate appropriate placeholder names such as pid, block\_id, ip, and port.
In the majority of responses, CodeLlama and CodeUp employ suitable placeholder names such as class\_name, method\_name, hostname, instance\_id, and version. 
However, sporadically, they resort to utilizing generic placeholders and irrelevant designations such as TPL, and MSG.
Notably, CodeLlama generated a generic template (i.e. \$DATE-\$TIME-...\$EVENT) in one instance, and in another, it omitted the template in the response while listing dynamic variables.
Llama 2 primarily parses log messages in accordance with a generic log message format (e.g. <timestamp> <class> <process\_id> <message\_body>), and occasionally, it produces generic names for variables such as \emph{value}, \emph{variable}, and \emph{var}.
Moreover, it occasionally generates irrelevant responses in HTML format.
Zephyr frequently opts for generic variable names such as dynamic\_variable\_1, VAR, TIMESTAMP, although it rarely assigns more appropriate names like NODE\_ID, APP\_ID.
Additionally, it generates timestamp formats in the YYYY-MM-dd HH:mm:ss structure uses [0-9] for integers, and employs placeholders such as XX, YY, \#\#\#, \_\_\_, $*$.
For 179 log messages, it responded with the same prompt provided as an input, and it did not send any responses for 9 log messages.

\head{Variable formatting}
Table~\ref{tab:variables} presents different variable formats generated in the responses by the LLMs in the zero-shot and few-shot experiments. 

GPT-3.5 mostly employs angle brackets to surround variables in accordance with the instructions provided in the prompt. 
However, it occasionally diverges from this convention by utilizing alternative formats such as \{\}, <TPL>, or employing opening and closing tags. 
Similarly, CodeUp primarily adopts angle brackets for variable formatting but may follow different formatting in some cases.
Claude 2.1 demonstrates a broader spectrum of bracketing styles and formatting techniques. 
While it commonly surrounds dynamic variables with angle brackets, it also incorporates other listed formats. 
Notably, there are instances where variables are omitted entirely without replacement with a placeholder.
CodeLlama mainly uses curly brackets to surround variables, whereas a minority adheres to the convention of angle brackets.
Rarely, it employs additional formatting styles outlined in Table~\ref{tab:variables} to surround the placeholders.

Llama 2 exhibits a wide range of bracketing styles and formatting variations for placeholders.
Notably, within a single log message, it may employ different formats for variables. 
Zephyr's behavior regarding variable formatting is characterized by instability and inconsistency, as Zephyr uses various formats for dynamic variables.
In some cases, the variable is either omitted entirely from the message or replaced with placeholders, such as "\_\_\_". 
In rare cases, it capitalizes the name of the variable without formatting any further. 
Furthermore, Zephyr occasionally diverges from a consistent formatting approach within a single log message, resulting in a mixture of various formats across different variables of the same log message. 
These non-deterministic tendencies present considerable challenges in discerning between templates and variables within responses, thus requiring strenuous effort to extract templates.

\begin{table*}
\centering
\caption{Variable formats followed by LLMs.}
\label{tab:variables}
\scalebox{0.9}{
\rowcolors{1}{}{lightgray}
\begin{tabular}{llllll}
\toprule
Model       & Zero-shot               & Few-shot         \\ \toprule
GPT-3.5     & \begin{tabular}[c]{@{}l@{}}$<>$, \{\}, X, <TPL>, \\ opening-closing tag pairs\end{tabular}      & $<*>$, $<>$           \\ %
Claude 2.1    & \begin{tabular}[c]{@{}l@{}}$<>$, \{\},  \$\{\}, @ @, \$ \$, \{\{\}\}, [[]], \\opening-closing tag pairs\end{tabular}  & $<*>$, $*$   \\ %
Llama 2    & \begin{tabular}[c]{@{}l@{}}\$\{\}, $<>$, \{\}, \$, \$ \$, [], []+, \$[], (), \%, ?, $**$ $**$, \#,  \\  no closing brackets, opening-closing tag pairs\end{tabular} & \begin{tabular}[c]{@{}l@{}} \{\}, $*$, $<>$, $<*>$, \\ $<*$ \end{tabular} \\ %
CodeLlama   & \begin{tabular}[c]{@{}l@{}}\{\}, $<>$, \$\{\}, \$, \%, \{\{\}\}, [], \\opening-closing tag pairs\end{tabular} & $<*>$, $<>$, $*$, $<* *>$    \\ %
Zephyr Beta & \begin{tabular}[c]{@{}l@{}} $<>$, \{\}, \{\{\}\}, [], [[]] \%, \#, \$, \$ \$, X  \\ \_\_\_, opening-closing tag pairs\end{tabular} & \begin{tabular}[c]{@{}l@{}} \{\}, ' ', [], \{\{\{\}\}\}, X \\ $*$, $<* *>$, $<*>$, $<*$ \end{tabular} \\ %
CodeUp      & \begin{tabular}[c]{@{}l@{}}$<>$, \{\}, \{\{\}\}, \%, \$\{\}, [], $<<>>$, $<>\%$, [\^ \_\^ ], \#, \& \\ no closing brackets, opening-closing tag pairs \end{tabular}  & $<*>$, $*$, $<*$, ... \\ \bottomrule
\end{tabular}
}
\end{table*}

\subsubsection{Observations from few-shot experiments} 
\label{subsection:observations_few-shot}
\noindent
\head{Placeholder selection}
In the majority of responses, GPT-3.5, Claude 2.1, CodeLlama, and CodeUp generate $*$ for the dynamic variables as shown in the examples of the few-shot prompt.
However, in rare cases, they produce variable names instead of using the $*$ sign. 
CodeUp exhibits a distinct behavior, where it removes only the header part (timestamp, etc.) from log messages and presents them as templates, particularly observed in the BLG and Mac projects. 

The Llama 2 responses with few-shot prompting are easier to use than their zero-shot counterparts.
It mostly uses a $*$ sign for variables, and variable names in some cases. 
However, it also produced a few plain messages without any templates and some templates with repetitive placeholders (i.e. $<*><*><*>....<*>$).

Zephyr returned the input prompt as a response for 364 log messages, while it yielded empty responses for 56 log messages. 
Occasionally, it generates generic templates, exemplified by structures such as \emph{<*CLASS*> <*LOGGER*> <*MESSAGE*> <*VARIABLE*>}.
Notably, responses generated for log messages originating from the BGL project were occasionally intertwined with HDFS log templates, a few of which were originally shared as examples in the prompt.
Throughout the remainder of its responses, it mostly uses $*$, and sometimes variable names or alternative designations like XXX or YYYY-MM-DD.

\head{Variable formatting}
GPT-3.5, Claude 2.1, CodeLlama, and CodeUp mostly generate angle brackets to surround the dynamic variables as shown in the examples of the few-shot prompt.
In some rare cases, they do not use brackets and just show $*$ or $<>$. %
In certain results from CodeUp, closing brackets are omitted for the variables (i.e. $<*$). 
While Llama and Zephyr primarily employ angle brackets for variable notation, they also incorporate different formats, as outlined in Table~\ref{tab:variables}. 

For an automated log parsing procedure, it is crucial that the template and variables can be easily parsed.
Therefore, we recommend the use of CodeLlama and CodeUp as free-to-use models when enough processing capabilities are available.
Otherwise, GPT3.5 creates results that are easy to parse and cheaper than Claude 2.1.

\begin{myframed}
\noindent
Llama 2 and Zephyr require a considerable amount of manual post-processing effort and are therefore not recommended. 
CodeLlama and CodeUp are recommended as free-to-use models, while GPT-3.5 is a cheaper alternative among the paid models.
\end{myframed}

\section{Concluding Remarks}
In this paper, we carried out an empirical study to compare the performance of six LLMs (2 paid, 4 free-to-use) for log parsing.
We evaluated their performance on three metrics and 1,354 log messages, each with a different underlying template.
Our findings show that among the six LLMs, CodeLlama and GPT3.5 perform the best.
CodeLlama is the best performing model for parsing accuracy and edit distance, two metrics applied in existing work.
When considering the longest common subsequence as a novel metric to determine the quality of the extracted templates based on string similarity, GPT-3.5 performs best.
While both ED and LCS are valid metrics that can be chosen based on the preferences/requirements for a given log parsing task (e.g., how important is the removal of timestamps), we are not able to recommend one over the other.
However, when it comes to the calculation of ED and LCS over a collection of log messages, we believe that the sum over normalized similarities is preferred, as this takes every message into account (unlike mean values) and outliers are scaled within the bounds of [0,1].

These results support future work in performing log parsing research without incurring costs (provided that hardware is available).
Practitioners can have high performance and increased privacy for log parsing by using CodeLlama.
Moreover, we highlight the importance of selecting the correct metrics for evaluating log parsing approaches, according to one's needs.

Another field of research that opens up is the combination of paid and free models. It can be interesting to determine the cases where free-to-use models are ``good enough'' to extract the template from a given log message and only use paid models for a fraction of log messages that were not parsed correctly.
Lastly, while we investigated two prompt types, further comparisons with different prompts and few-shot examples could be conducted to improve the performance of LLMs.

\section*{Data Availability}

\noindent
To support open science and allow for replication and verification of our work, 
all artifacts are made available through Zenodo at the following URL:
{\url{https://doi.org/10.5281/zenodo.13625383}.
}

\begin{acks}
This work is supported by the Research Council of Norway through the cureIT (IKTPLUSS \#300461) and the secureIT projects (IKTPLUSS \#288787), and by the European Union through the Horizon Europe Marie Sk\l{}odowska-Curie Actions (\#101151798).
The empirical evaluation was performed on the Experimental Infrastructure for Exploration of Exascale Computing (eX3), 
financially supported by the Research Council of Norway under contract \#270053.
\end{acks}
\bibliographystyle{ACM-Reference-Format}
\input{arXiv_2024_log_parsing.bbl}

 \end{document}

%% file: arXiv_2024_log_parsing.bbl

%% file: arXiv_2024_log_parsing.bbl
\begin{thebibliography}{32}


\ifx \showCODEN    \undefined \def \showCODEN     #1{\unskip}     \fi
\ifx \showDOI      \undefined \def \showDOI       #1{#1}\fi
\ifx \showISBNx    \undefined \def \showISBNx     #1{\unskip}     \fi
\ifx \showISBNxiii \undefined \def \showISBNxiii  #1{\unskip}     \fi
\ifx \showISSN     \undefined \def \showISSN      #1{\unskip}     \fi
\ifx \showLCCN     \undefined \def \showLCCN      #1{\unskip}     \fi
\ifx \shownote     \undefined \def \shownote      #1{#1}          \fi
\ifx \showarticletitle \undefined \def \showarticletitle #1{#1}   \fi
\ifx \showURL      \undefined \def \showURL       {\relax}        \fi
\providecommand\bibfield[2]{#2}
\providecommand\bibinfo[2]{#2}
\providecommand\natexlab[1]{#1}
\providecommand\showeprint[2][]{arXiv:#2}

\bibitem[Astekin et~al\mbox{.}(2024)]%
        {astekin2024exploratory}
\bibfield{author}{\bibinfo{person}{Merve Astekin}, \bibinfo{person}{Max Hort},
  {and} \bibinfo{person}{Leon Moonen}.} \bibinfo{year}{2024}\natexlab{}.
\newblock \showarticletitle{An Exploratory Study on How Non-Determinism in
  Large Language Models Affects Log Parsing}. In
  \bibinfo{booktitle}{\emph{Proceedings of the ACM/IEEE 2nd International
  Workshop on Interpretability, Robustness, and Benchmarking in Neural Software
  Engineering}}. \bibinfo{pages}{13--18}.
\newblock


\bibitem[Astekin et~al\mbox{.}(2018)]%
        {astekin2019:dilaf}
\bibfield{author}{\bibinfo{person}{Merve Astekin}, \bibinfo{person}{Harun
  Zengin}, {and} \bibinfo{person}{Hasan S{\"o}zer}.}
  \bibinfo{year}{2018}\natexlab{}.
\newblock \showarticletitle{DILAF: A framework for distributed analysis of
  large‐scale system logs for anomaly detection}.
\newblock \bibinfo{journal}{\emph{Software: Practice and Experience}}
  \bibinfo{volume}{49} (\bibinfo{year}{2018}), \bibinfo{pages}{153 -- 170}.
\newblock
\urldef\tempurl%
\url{https://api.semanticscholar.org/CorpusID:59222641}
\showURL{%
\tempurl}


\bibitem[Chen and Liao(2022)]%
        {chen2022:bertlog}
\bibfield{author}{\bibinfo{person}{Song Chen} {and} \bibinfo{person}{Hai
  Liao}.} \bibinfo{year}{2022}\natexlab{}.
\newblock \showarticletitle{{{BERT-Log}}: {{Anomaly Detection}} for {{System
  Logs Based}} on {{Pre-trained Language Model}}}.
\newblock \bibinfo{journal}{\emph{Applied Artificial Intelligence}}
  \bibinfo{volume}{36}, \bibinfo{number}{1} (\bibinfo{date}{Dec.}
  \bibinfo{year}{2022}), \bibinfo{pages}{2145642}.
\newblock
\showISSN{0883-9514}
\urldef\tempurl%
\url{https://doi.org/10.1080/08839514.2022.2145642}
\showDOI{\tempurl}


\bibitem[Dai et~al\mbox{.}(2022)]%
        {dai2022:logram}
\bibfield{author}{\bibinfo{person}{Hetong Dai}, \bibinfo{person}{Heng Li},
  \bibinfo{person}{Che-Shao Chen}, \bibinfo{person}{Weiyi Shang}, {and}
  \bibinfo{person}{Tse-Hsun Chen}.} \bibinfo{year}{2022}\natexlab{}.
\newblock \showarticletitle{Logram: {{Efficient Log Parsing Using}} Nn-{{Gram
  Dictionaries}}}.
\newblock \bibinfo{journal}{\emph{IEEE Transactions on Software Engineering}}
  \bibinfo{volume}{48}, \bibinfo{number}{3} (\bibinfo{date}{March}
  \bibinfo{year}{2022}), \bibinfo{pages}{879--892}.
\newblock
\showISSN{1939-3520}
\urldef\tempurl%
\url{https://doi.org/10.1109/TSE.2020.3007554}
\showDOI{\tempurl}


\bibitem[Dai et~al\mbox{.}(2020)]%
        {dai2020:logram}
\bibfield{author}{\bibinfo{person}{Hetong Dai}, \bibinfo{person}{Heng Li},
  \bibinfo{person}{Weiyi Shang}, \bibinfo{person}{Tse-Hsun Chen}, {and}
  \bibinfo{person}{Che-Shao Chen}.} \bibinfo{year}{2020}\natexlab{}.
\newblock \bibinfo{title}{Logram: {{Efficient Log Parsing Using}} n-{{Gram
  Dictionaries}}}.
\newblock
\newblock
\showeprint[arxiv]{2001.03038}~[cs]


\bibitem[Du and Li(2016)]%
        {du2016:spell}
\bibfield{author}{\bibinfo{person}{Min Du} {and} \bibinfo{person}{Feifei Li}.}
  \bibinfo{year}{2016}\natexlab{}.
\newblock \showarticletitle{Spell: {{Streaming Parsing}} of {{System Event
  Logs}}}. In \bibinfo{booktitle}{\emph{2016 {{IEEE}} 16th {{International
  Conference}} on {{Data Mining}} ({{ICDM}})}}. \bibinfo{pages}{859--864}.
\newblock
\showISSN{2374-8486}
\urldef\tempurl%
\url{https://doi.org/10.1109/ICDM.2016.0103}
\showDOI{\tempurl}


\bibitem[Guo et~al\mbox{.}(2021)]%
        {guo2021:LogBERT}
\bibfield{author}{\bibinfo{person}{Haixuan Guo}, \bibinfo{person}{Shuhan Yuan},
  {and} \bibinfo{person}{Xintao Wu}.} \bibinfo{year}{2021}\natexlab{}.
\newblock \showarticletitle{{{LogBERT}}: {{Log Anomaly Detection}} via
  {{BERT}}}. In \bibinfo{booktitle}{\emph{2021 {{International Joint
  Conference}} on {{Neural Networks}} ({{IJCNN}})}}. \bibinfo{pages}{1--8}.
\newblock
\showISSN{2161-4407}
\urldef\tempurl%
\url{https://doi.org/10.1109/IJCNN52387.2021.9534113}
\showDOI{\tempurl}


\bibitem[He et~al\mbox{.}(2017)]%
        {he2017:drain}
\bibfield{author}{\bibinfo{person}{Pinjia He}, \bibinfo{person}{Jieming Zhu},
  \bibinfo{person}{Zibin Zheng}, {and} \bibinfo{person}{Michael~R. Lyu}.}
  \bibinfo{year}{2017}\natexlab{}.
\newblock \showarticletitle{Drain: {{An Online Log Parsing Approach}} with
  {{Fixed Depth Tree}}}. In \bibinfo{booktitle}{\emph{International
  {{Conference}} on {{Web Services}} ({{ICWS}})}}. \bibinfo{publisher}{IEEE},
  \bibinfo{pages}{33--40}.
\newblock
\showISBNx{978-1-5386-0752-7}
\urldef\tempurl%
\url{https://doi.org/10.1109/icws.2017.13}
\showDOI{\tempurl}


\bibitem[Huo et~al\mbox{.}(2023)]%
        {huo2023:semparser}
\bibfield{author}{\bibinfo{person}{Yintong Huo}, \bibinfo{person}{Yuxin Su},
  \bibinfo{person}{Cheryl Lee}, {and} \bibinfo{person}{Michael~R. Lyu}.}
  \bibinfo{year}{2023}\natexlab{}.
\newblock \bibinfo{title}{{{SemParser}}: {{A Semantic Parser}} for {{Log
  Analysis}}}.
\newblock
\newblock
\showeprint[arxiv]{2112.12636}~[cs]


\bibitem[Jiang and Kim(2023)]%
        {jiang2023:codeup}
\bibfield{author}{\bibinfo{person}{Juyong Jiang} {and} \bibinfo{person}{Sunghun
  Kim}.} \bibinfo{year}{2023}\natexlab{}.
\newblock \bibinfo{title}{{{CodeUp}}: {{A Multilingual Code Generation Llama2
  Model}} with {{Parameter-Efficient Instruction-Tuning}}}.
\newblock
\newblock


\bibitem[Jiang et~al\mbox{.}(2024a)]%
        {jiang2024:lilac}
\bibfield{author}{\bibinfo{person}{Zhihan Jiang}, \bibinfo{person}{Jinyang
  Liu}, \bibinfo{person}{Zhuangbin Chen}, \bibinfo{person}{Yichen Li},
  \bibinfo{person}{Junjie Huang}, \bibinfo{person}{Yintong Huo},
  \bibinfo{person}{Pinjia He}, \bibinfo{person}{Jiazhen Gu}, {and}
  \bibinfo{person}{Michael~R. Lyu}.} \bibinfo{year}{2024}\natexlab{a}.
\newblock \bibinfo{title}{{{LILAC}}: {{Log Parsing}} Using {{LLMs}} with
  {{Adaptive Parsing Cache}}}.
\newblock
\newblock
\showeprint[arxiv]{2310.01796}~[cs]


\bibitem[Jiang et~al\mbox{.}(2024b)]%
        {jiang2024:largescale}
\bibfield{author}{\bibinfo{person}{Zhihan Jiang}, \bibinfo{person}{Jinyang
  Liu}, \bibinfo{person}{Junjie Huang}, \bibinfo{person}{Yichen Li},
  \bibinfo{person}{Yintong Huo}, \bibinfo{person}{Jiazhen Gu},
  \bibinfo{person}{Zhuangbin Chen}, \bibinfo{person}{Jieming Zhu}, {and}
  \bibinfo{person}{Michael~R. Lyu}.} \bibinfo{year}{2024}\natexlab{b}.
\newblock \bibinfo{title}{A {{Large-Scale Evaluation}} for {{Log Parsing
  Techniques}}: {{How Far Are We}}?}
\newblock
\newblock
\urldef\tempurl%
\url{https://doi.org/10.48550/arXiv.2308.10828}
\showDOI{\tempurl}
\showeprint[arxiv]{2308.10828}~[cs]


\bibitem[Khan et~al\mbox{.}(2022)]%
        {khan2022:guidelines}
\bibfield{author}{\bibinfo{person}{Zanis~Ali Khan}, \bibinfo{person}{Donghwan
  Shin}, \bibinfo{person}{Domenico Bianculli}, {and} \bibinfo{person}{Lionel
  Briand}.} \bibinfo{year}{2022}\natexlab{}.
\newblock \showarticletitle{Guidelines for Assessing the Accuracy of Log
  Message Template Identification Techniques}. In
  \bibinfo{booktitle}{\emph{Proceedings of the 44th {{International
  Conference}} on {{Software Engineering}}}}. \bibinfo{publisher}{ACM},
  \bibinfo{address}{Pittsburgh Pennsylvania}, \bibinfo{pages}{1095--1106}.
\newblock
\showISBNx{978-1-4503-9221-1}
\urldef\tempurl%
\url{https://doi.org/10.1145/3510003.3510101}
\showDOI{\tempurl}


\bibitem[Le and Zhang(2022)]%
        {le2022:logbased}
\bibfield{author}{\bibinfo{person}{Van-Hoang Le} {and} \bibinfo{person}{Hongyu
  Zhang}.} \bibinfo{year}{2022}\natexlab{}.
\newblock \showarticletitle{Log-Based Anomaly Detection with Deep Learning: How
  Far Are We?}. In \bibinfo{booktitle}{\emph{Proceedings of the 44th
  {{International Conference}} on {{Software Engineering}}}}
  \emph{(\bibinfo{series}{{{ICSE}} '22})}. \bibinfo{publisher}{Association for
  Computing Machinery}, \bibinfo{address}{New York, NY, USA},
  \bibinfo{pages}{1356--1367}.
\newblock
\showISBNx{978-1-4503-9221-1}
\urldef\tempurl%
\url{https://doi.org/10.1145/3510003.3510155}
\showDOI{\tempurl}


\bibitem[Le and Zhang(2023a)]%
        {le2023:loga}
\bibfield{author}{\bibinfo{person}{Van-Hoang Le} {and} \bibinfo{person}{Hongyu
  Zhang}.} \bibinfo{year}{2023}\natexlab{a}.
\newblock \showarticletitle{Log {{Parsing}}: {{How Far Can ChatGPT Go}}?}. In
  \bibinfo{booktitle}{\emph{2023 38th {{IEEE}}/{{ACM International Conference}}
  on {{Automated Software Engineering}} ({{ASE}})}}. \bibinfo{publisher}{IEEE},
  \bibinfo{address}{Luxembourg, Luxembourg}, \bibinfo{pages}{1699--1704}.
\newblock
\showISBNx{9798350329964}
\urldef\tempurl%
\url{https://doi.org/10.1109/ase56229.2023.00206}
\showDOI{\tempurl}


\bibitem[Le and Zhang(2023b)]%
        {le2023:log}
\bibfield{author}{\bibinfo{person}{Van-Hoang Le} {and} \bibinfo{person}{Hongyu
  Zhang}.} \bibinfo{year}{2023}\natexlab{b}.
\newblock \showarticletitle{Log {{Parsing}} with {{Prompt-Based Few-Shot
  Learning}}}. In \bibinfo{booktitle}{\emph{Proceedings of the 45th
  {{International Conference}} on {{Software Engineering}}}}
  \emph{(\bibinfo{series}{{{ICSE}} '23})}. \bibinfo{publisher}{IEEE Press},
  \bibinfo{address}{Melbourne, Victoria, Australia},
  \bibinfo{pages}{2438--2449}.
\newblock
\showISBNx{978-1-66545-701-9}
\urldef\tempurl%
\url{https://doi.org/10.1109/icse48619.2023.00204}
\showDOI{\tempurl}


\bibitem[Lee et~al\mbox{.}(2023)]%
        {lee2023:lanobert}
\bibfield{author}{\bibinfo{person}{Yukyung Lee}, \bibinfo{person}{Jina Kim},
  {and} \bibinfo{person}{Pilsung Kang}.} \bibinfo{year}{2023}\natexlab{}.
\newblock \showarticletitle{{{LAnoBERT}}: {{System}} Log Anomaly Detection
  Based on {{BERT}} Masked Language Model}.
\newblock \bibinfo{journal}{\emph{Applied Soft Computing}}
  \bibinfo{volume}{146} (\bibinfo{date}{Oct.} \bibinfo{year}{2023}),
  \bibinfo{pages}{110689}.
\newblock
\showISSN{1568-4946}
\urldef\tempurl%
\url{https://doi.org/10.1016/j.asoc.2023.110689}
\showDOI{\tempurl}


\bibitem[Li et~al\mbox{.}(2023)]%
        {li2023:did}
\bibfield{author}{\bibinfo{person}{Zhenhao Li}, \bibinfo{person}{Chuan Luo},
  \bibinfo{person}{Tse-Hsun~(Peter) Chen}, \bibinfo{person}{Weiyi Shang},
  \bibinfo{person}{Shilin He}, \bibinfo{person}{Qingwei Lin}, {and}
  \bibinfo{person}{Dongmei Zhang}.} \bibinfo{year}{2023}\natexlab{}.
\newblock \showarticletitle{Did {{We Miss Something Important}}? {{Studying}}
  and {{Exploring Variable-Aware Log Abstraction}}}. In
  \bibinfo{booktitle}{\emph{Proceedings of the 45th {{International
  Conference}} on {{Software Engineering}}}} \emph{(\bibinfo{series}{{{ICSE}}
  '23})}. \bibinfo{publisher}{IEEE Press}, \bibinfo{address}{Melbourne,
  Victoria, Australia}, \bibinfo{pages}{830--842}.
\newblock
\showISBNx{978-1-66545-701-9}
\urldef\tempurl%
\url{https://doi.org/10.1109/ICSE48619.2023.00078}
\showDOI{\tempurl}


\bibitem[Liu et~al\mbox{.}(2023)]%
        {liu2023:logprompt}
\bibfield{author}{\bibinfo{person}{Yilun Liu}, \bibinfo{person}{Shimin Tao},
  \bibinfo{person}{Weibin Meng}, \bibinfo{person}{Jingyu Wang},
  \bibinfo{person}{Wenbing Ma}, \bibinfo{person}{Yanqing Zhao},
  \bibinfo{person}{Yuhang Chen}, \bibinfo{person}{Hao Yang},
  \bibinfo{person}{Yanfei Jiang}, {and} \bibinfo{person}{Xun Chen}.}
  \bibinfo{year}{2023}\natexlab{}.
\newblock \bibinfo{title}{{{LogPrompt}}: {{Prompt Engineering Towards
  Zero-Shot}} and {{Interpretable Log Analysis}}}.
\newblock
\newblock
\urldef\tempurl%
\url{https://doi.org/10.48550/arXiv.2308.07610}
\showDOI{\tempurl}
\showeprint[arxiv]{2308.07610}~[cs]


\bibitem[Liu et~al\mbox{.}(2024)]%
        {liu2024:interpretable}
\bibfield{author}{\bibinfo{person}{Yilun Liu}, \bibinfo{person}{Shimin Tao},
  \bibinfo{person}{Weibin Meng}, \bibinfo{person}{Jingyu Wang},
  \bibinfo{person}{Wenbing Ma}, \bibinfo{person}{Yanqing Zhao},
  \bibinfo{person}{Yuhang Chen}, \bibinfo{person}{Hao Yang},
  \bibinfo{person}{Yanfei Jiang}, {and} \bibinfo{person}{Xun Chen}.}
  \bibinfo{year}{2024}\natexlab{}.
\newblock \bibinfo{title}{Interpretable {{Online Log Analysis Using Large
  Language Models}} with {{Prompt Strategies}}}.
\newblock
\newblock
\urldef\tempurl%
\url{https://doi.org/10.48550/arXiv.2308.07610}
\showDOI{\tempurl}
\showeprint[arxiv]{2308.07610}~[cs]


\bibitem[Liu et~al\mbox{.}(2022)]%
        {liu2022:uniparser}
\bibfield{author}{\bibinfo{person}{Yudong Liu}, \bibinfo{person}{Xu Zhang},
  \bibinfo{person}{Shilin He}, \bibinfo{person}{Hongyu Zhang},
  \bibinfo{person}{Liqun Li}, \bibinfo{person}{Yu Kang}, \bibinfo{person}{Yong
  Xu}, \bibinfo{person}{Minghua Ma}, \bibinfo{person}{Qingwei Lin},
  \bibinfo{person}{Yingnong Dang}, \bibinfo{person}{Saravan Rajmohan}, {and}
  \bibinfo{person}{Dongmei Zhang}.} \bibinfo{year}{2022}\natexlab{}.
\newblock \showarticletitle{{{UniParser}}: {{A Unified Log Parser}} for
  {{Heterogeneous Log Data}}}. In \bibinfo{booktitle}{\emph{Proceedings of the
  {{ACM Web Conference}} 2022}}. \bibinfo{pages}{1893--1901}.
\newblock
\urldef\tempurl%
\url{https://doi.org/10.1145/3485447.3511993}
\showDOI{\tempurl}
\showeprint[arxiv]{2202.06569}~[cs]


\bibitem[Ma et~al\mbox{.}(2024)]%
        {ma2024:llmparser}
\bibfield{author}{\bibinfo{person}{Zeyang Ma}, \bibinfo{person}{An~Ran Chen},
  \bibinfo{person}{Dong~Jae Kim}, \bibinfo{person}{Tse-Hsun Chen}, {and}
  \bibinfo{person}{Shaowei Wang}.} \bibinfo{year}{2024}\natexlab{}.
\newblock \showarticletitle{{{LLMParser}}: {{An Exploratory Study}} on {{Using
  Large Language Models}} for {{Log Parsing}}}. In
  \bibinfo{booktitle}{\emph{Proceedings of the {{IEEE}}/{{ACM}} 46th
  {{International Conference}} on {{Software Engineering}}}}
  \emph{(\bibinfo{series}{{{ICSE}} '24})}. \bibinfo{publisher}{Association for
  Computing Machinery}, \bibinfo{address}{New York, NY, USA},
  \bibinfo{pages}{1--13}.
\newblock
\showISBNx{9798400702174}
\urldef\tempurl%
\url{https://doi.org/10.1145/3597503.3639150}
\showDOI{\tempurl}


\bibitem[Mudgal and Wouhaybi(2023)]%
        {mudgal2023:assessment}
\bibfield{author}{\bibinfo{person}{Priyanka Mudgal} {and} \bibinfo{person}{Rita
  Wouhaybi}.} \bibinfo{year}{2023}\natexlab{}.
\newblock \bibinfo{title}{An {{Assessment}} of {{ChatGPT}} on {{Log Data}}}.
\newblock
\newblock
\urldef\tempurl%
\url{https://doi.org/10.48550/arXiv.2309.07938}
\showDOI{\tempurl}
\showeprint[arxiv]{2309.07938}~[cs]


\bibitem[Nedelkoski et~al\mbox{.}(2021)]%
        {nedelkoski2021:selfsuperviseda}
\bibfield{author}{\bibinfo{person}{Sasho Nedelkoski}, \bibinfo{person}{Jasmin
  Bogatinovski}, \bibinfo{person}{Alexander Acker}, \bibinfo{person}{Jorge
  Cardoso}, {and} \bibinfo{person}{Odej Kao}.} \bibinfo{year}{2021}\natexlab{}.
\newblock \showarticletitle{Self-Supervised {{Log Parsing}}}.
\newblock \bibinfo{journal}{\emph{Machine Learning and Knowledge Discovery in
  Databases: Applied Data Science Track}}  \bibinfo{volume}{12460}
  (\bibinfo{year}{2021}), \bibinfo{pages}{122--138}.
\newblock
\showISBNx{9783030676667 9783030676674}
\urldef\tempurl%
\url{https://doi.org/10.1007/978-3-030-67667-4_8}
\showDOI{\tempurl}


\bibitem[Petrescu et~al\mbox{.}(2023)]%
        {petrescu2023:log}
\bibfield{author}{\bibinfo{person}{Stefan Petrescu}, \bibinfo{person}{Floris
  den Hengst}, \bibinfo{person}{Alexandru Uta}, {and} \bibinfo{person}{Jan~S.
  Rellermeyer}.} \bibinfo{year}{2023}\natexlab{}.
\newblock \bibinfo{title}{Log {{Parsing Evaluation}} in the {{Era}} of {{Modern
  Software Systems}}}.
\newblock
\newblock
\showeprint[arxiv]{2308.09003}~[cs]


\bibitem[Qi et~al\mbox{.}(2023)]%
        {qi2023:loggpt}
\bibfield{author}{\bibinfo{person}{Jiaxing Qi}, \bibinfo{person}{Shaohan
  Huang}, \bibinfo{person}{Zhongzhi Luan}, \bibinfo{person}{Carol Fung},
  \bibinfo{person}{Hailong Yang}, {and} \bibinfo{person}{Depei Qian}.}
  \bibinfo{year}{2023}\natexlab{}.
\newblock \bibinfo{title}{{{LogGPT}}: {{Exploring ChatGPT}} for {{Log-Based
  Anomaly Detection}}}.
\newblock
\newblock
\urldef\tempurl%
\url{https://doi.org/10.48550/arXiv.2309.01189}
\showDOI{\tempurl}
\showeprint[arxiv]{2309.01189}~[cs]


\bibitem[Rozi{\`e}re et~al\mbox{.}(2023)]%
        {roziere2023:code}
\bibfield{author}{\bibinfo{person}{Baptiste Rozi{\`e}re},
  \bibinfo{person}{Jonas Gehring}, \bibinfo{person}{Fabian Gloeckle},
  \bibinfo{person}{Sten Sootla}, \bibinfo{person}{Itai Gat},
  \bibinfo{person}{Xiaoqing~Ellen Tan}, \bibinfo{person}{Yossi Adi},
  \bibinfo{person}{Jingyu Liu}, \bibinfo{person}{Tal Remez},
  \bibinfo{person}{J{\'e}r{\'e}my Rapin}, \bibinfo{person}{Artyom Kozhevnikov},
  \bibinfo{person}{Ivan Evtimov}, \bibinfo{person}{Joanna Bitton},
  \bibinfo{person}{Manish Bhatt}, \bibinfo{person}{Cristian~Canton Ferrer},
  \bibinfo{person}{Aaron Grattafiori}, \bibinfo{person}{Wenhan Xiong},
  \bibinfo{person}{Alexandre D{\'e}fossez}, \bibinfo{person}{Jade Copet},
  \bibinfo{person}{Faisal Azhar}, \bibinfo{person}{Hugo Touvron},
  \bibinfo{person}{Louis Martin}, \bibinfo{person}{Nicolas Usunier},
  \bibinfo{person}{Thomas Scialom}, {and} \bibinfo{person}{Gabriel Synnaeve}.}
  \bibinfo{year}{2023}\natexlab{}.
\newblock \bibinfo{title}{Code {{Llama}}: {{Open Foundation Models}} for
  {{Code}}}.
\newblock
\newblock
\showeprint[arxiv]{2308.12950}~[cs]


\bibitem[Touvron et~al\mbox{.}(2023)]%
        {touvron2023:llama}
\bibfield{author}{\bibinfo{person}{Hugo Touvron}, \bibinfo{person}{Louis
  Martin}, \bibinfo{person}{Kevin Stone}, \bibinfo{person}{Peter Albert},
  \bibinfo{person}{Amjad Almahairi}, \bibinfo{person}{Yasmine Babaei},
  \bibinfo{person}{Nikolay Bashlykov}, \bibinfo{person}{Soumya Batra},
  \bibinfo{person}{Prajjwal Bhargava}, \bibinfo{person}{Shruti Bhosale},
  \bibinfo{person}{Dan Bikel}, \bibinfo{person}{Lukas Blecher},
  \bibinfo{person}{Cristian~Canton Ferrer}, \bibinfo{person}{Moya Chen},
  \bibinfo{person}{Guillem Cucurull}, \bibinfo{person}{David Esiobu},
  \bibinfo{person}{Jude Fernandes}, \bibinfo{person}{Jeremy Fu},
  \bibinfo{person}{Wenyin Fu}, \bibinfo{person}{Brian Fuller},
  \bibinfo{person}{Cynthia Gao}, \bibinfo{person}{Vedanuj Goswami},
  \bibinfo{person}{Naman Goyal}, \bibinfo{person}{Anthony Hartshorn},
  \bibinfo{person}{Saghar Hosseini}, \bibinfo{person}{Rui Hou},
  \bibinfo{person}{Hakan Inan}, \bibinfo{person}{Marcin Kardas},
  \bibinfo{person}{Viktor Kerkez}, \bibinfo{person}{Madian Khabsa},
  \bibinfo{person}{Isabel Kloumann}, \bibinfo{person}{Artem Korenev},
  \bibinfo{person}{Punit~Singh Koura}, \bibinfo{person}{Marie-Anne Lachaux},
  \bibinfo{person}{Thibaut Lavril}, \bibinfo{person}{Jenya Lee},
  \bibinfo{person}{Diana Liskovich}, \bibinfo{person}{Yinghai Lu},
  \bibinfo{person}{Yuning Mao}, \bibinfo{person}{Xavier Martinet},
  \bibinfo{person}{Todor Mihaylov}, \bibinfo{person}{Pushkar Mishra},
  \bibinfo{person}{Igor Molybog}, \bibinfo{person}{Yixin Nie},
  \bibinfo{person}{Andrew Poulton}, \bibinfo{person}{Jeremy Reizenstein},
  \bibinfo{person}{Rashi Rungta}, \bibinfo{person}{Kalyan Saladi},
  \bibinfo{person}{Alan Schelten}, \bibinfo{person}{Ruan Silva},
  \bibinfo{person}{Eric~Michael Smith}, \bibinfo{person}{Ranjan Subramanian},
  \bibinfo{person}{Xiaoqing~Ellen Tan}, \bibinfo{person}{Binh Tang},
  \bibinfo{person}{Ross Taylor}, \bibinfo{person}{Adina Williams},
  \bibinfo{person}{Jian~Xiang Kuan}, \bibinfo{person}{Puxin Xu},
  \bibinfo{person}{Zheng Yan}, \bibinfo{person}{Iliyan Zarov},
  \bibinfo{person}{Yuchen Zhang}, \bibinfo{person}{Angela Fan},
  \bibinfo{person}{Melanie Kambadur}, \bibinfo{person}{Sharan Narang},
  \bibinfo{person}{Aurelien Rodriguez}, \bibinfo{person}{Robert Stojnic},
  \bibinfo{person}{Sergey Edunov}, {and} \bibinfo{person}{Thomas Scialom}.}
  \bibinfo{year}{2023}\natexlab{}.
\newblock \bibinfo{title}{Llama 2: {{Open Foundation}} and {{Fine-Tuned Chat
  Models}}}.
\newblock
\newblock
\showeprint[arxiv]{2307.09288}~[cs]


\bibitem[Tunstall et~al\mbox{.}(2023)]%
        {tunstall2023:zephyr}
\bibfield{author}{\bibinfo{person}{Lewis Tunstall}, \bibinfo{person}{Edward
  Beeching}, \bibinfo{person}{Nathan Lambert}, \bibinfo{person}{Nazneen
  Rajani}, \bibinfo{person}{Kashif Rasul}, \bibinfo{person}{Younes Belkada},
  \bibinfo{person}{Shengyi Huang}, \bibinfo{person}{Leandro {von Werra}},
  \bibinfo{person}{Cl{\'e}mentine Fourrier}, \bibinfo{person}{Nathan Habib},
  \bibinfo{person}{Nathan Sarrazin}, \bibinfo{person}{Omar Sanseviero},
  \bibinfo{person}{Alexander~M. Rush}, {and} \bibinfo{person}{Thomas Wolf}.}
  \bibinfo{year}{2023}\natexlab{}.
\newblock \bibinfo{title}{Zephyr: {{Direct Distillation}} of {{LM Alignment}}}.
\newblock
\newblock
\showeprint[arxiv]{2310.16944}~[cs]


\bibitem[Xu et~al\mbox{.}(2024)]%
        {xu2024:divlog}
\bibfield{author}{\bibinfo{person}{Junjielong Xu}, \bibinfo{person}{Ruichun
  Yang}, \bibinfo{person}{Yintong Huo}, \bibinfo{person}{Chengyu Zhang}, {and}
  \bibinfo{person}{Pinjia He}.} \bibinfo{year}{2024}\natexlab{}.
\newblock \showarticletitle{{{DivLog}}: {{Log Parsing}} with {{Prompt Enhanced
  In-Context Learning}}}. In \bibinfo{booktitle}{\emph{Proceedings of the
  {{IEEE}}/{{ACM}} 46th {{International Conference}} on {{Software
  Engineering}}}}. \bibinfo{publisher}{ACM}, \bibinfo{address}{Lisbon
  Portugal}, \bibinfo{pages}{1--12}.
\newblock
\showISBNx{9798400702174}
\urldef\tempurl%
\url{https://doi.org/10.1145/3597503.3639155}
\showDOI{\tempurl}


\bibitem[Zhu et~al\mbox{.}(2023)]%
        {zhu2023:loghub}
\bibfield{author}{\bibinfo{person}{Jieming Zhu}, \bibinfo{person}{Shilin He},
  \bibinfo{person}{Pinjia He}, \bibinfo{person}{Jinyang Liu}, {and}
  \bibinfo{person}{Michael~R. Lyu}.} \bibinfo{year}{2023}\natexlab{}.
\newblock \bibinfo{title}{Loghub: {{A Large Collection}} of {{System Log
  Datasets}} for {{AI-driven Log Analytics}}}.
\newblock
\newblock
\showeprint[arxiv]{2008.06448}~[cs]


\bibitem[Zhu et~al\mbox{.}(2019)]%
        {zhu2019:tools}
\bibfield{author}{\bibinfo{person}{Jieming Zhu}, \bibinfo{person}{Shilin He},
  \bibinfo{person}{Jinyang Liu}, \bibinfo{person}{Pinjia He},
  \bibinfo{person}{Qi Xie}, \bibinfo{person}{Zibin Zheng}, {and}
  \bibinfo{person}{Michael~R. Lyu}.} \bibinfo{year}{2019}\natexlab{}.
\newblock \showarticletitle{Tools and {{Benchmarks}} for {{Automated Log
  Parsing}}}. In \bibinfo{booktitle}{\emph{{{ICSE}}: {{Software Engineering}}
  in {{Practice}} ({{ICSE-SEIP}})}}. \bibinfo{publisher}{IEEE},
  \bibinfo{address}{Montreal, QC, Canada}, \bibinfo{pages}{121--130}.
\newblock
\showISBNx{978-1-72811-760-7}
\urldef\tempurl%
\url{https://doi.org/10.1109/icse-seip.2019.00021}
\showDOI{\tempurl}


\end{thebibliography}
